\documentclass[epj,nopacs]{svjour}
\usepackage{graphics}
\usepackage{amsmath}
\usepackage{multirow}
\usepackage{color}

\begin{document}
\title{A scheme to fix multiple solutions in amplitude analyses}
\author{Yuanning Gao\inst{1}, Tianze Rong\inst{1}, Zhenwei Yang\inst{1}, Chenjia Zhang\inst{1}, Yanxi Zhang\inst{1}
}                     
%
%
\institute{School of Physics, Peking University, Beijing 100871, China}
\date{}
%
\abstract{
Decays of unstable heavy particles usually involve the coherent sum of several amplitudes, like in a multiple slit experiment.  Dedicated amplitude analysis techniques have been widely used to resolve these amplitudes for a better understanding of the underlying dynamics. For special cases, where two spin-1/2 particles and two (pseudo-)scalar particles are present in the process, multiple equivalent solutions are found due to intrinsic symmetries in the summed probability density function.
In this paper, the problem of multiple solutions is discussed and a scheme to overcome this problem is proposed by fixing some free parameters. Toys are generated to validate the strategy. A new approach to align helicities of initial- and final-state particles in different decay chains is also introduced.
\PACS{
      {PACS-key}{discribing text of that key}   \and
      {PACS-key}{discribing text of that key}
     } 
} 
\authorrunning{Y. Gao et al.}
\maketitle

\section{Introduction}
\label{section:Introduction}
Multi-body decays of unstable particles provide richer experimental information compared to two-body decays due to the involvement of various intermediate resonances. It makes multi-body decays extremely important in the test of the Standard Model of particle physics and search for new resonances.
As a few examples, 
the $H^0\to Z\mu^+\mu^-$ decay is used to determine the Higgs spin~\cite{CMS:2012vby}, the $\Lambda_b^0\to J/\psi p K^-$ decay gives rises to the first observation of pentaquark states~\cite{LHCb:2015yax}, 
and recently the LHCb collaboration observes various sources of violations of  charge-conjugation and parity asymmetries in the final-state phase space of $B$ meson decays into three hadrons~\cite{LHCb:2022nyw}. 
Following the isobar model~\cite{Fleming:1964zz,osti_4551044,PhysRevD.11.3165}, the total amplitude of a multi-body decay of a hadron can be written as the coherent sum of several sub-amplitudes, 
each one with a definite helicity for all participating particles. 
The square of the total amplitude modulus, summed over the initial and final state helicities, gives the final-state probability density distribution (PDF). 
Unknown parameters, such as the contribution of a new resonance in the decay, the spin-parity $J^\eta$, and the strength of a specific amplitude, can be extracted by fitting the PDF to real data.
Three-body decays $P_0\rightarrow P_1P_2P_3$ that involve two spin-$1/2$ fermions and two (pseudo-)scalars (denoted as $h$ in the following) are often studied in heavy flavor physics. They are referred to as \texttt{2F2P} decays in the following. They include the $\Lambda_b^0\rightarrow D^0p h^- $~\cite{LHCb:2017jym,Zhang:2021sit},
$\Xi_b^-\to pK^-K^-$~\cite{LHCb:2021enr},
$\Lambda_c^+\rightarrow pK^-\pi^+$~\cite{LHCb:2022ouv} and $B\rightarrow p \overline{p}^- h$ decays {\em etc.}. When fitting to experimental data, the PDF of such decays has symmetries that prevent a unique determination of all free parameters, namely there are multiple equivalent solutions. In this article, this problem is demonstrated and a possible strategy to fix multiple solutions is proposed.

\section{Helicity amplitude}
\label{section:HelAmp}
Dedicated approaches based on the fundamental symmetry of Lorentz invariance have been adopted to prepare the amplitudes of a multi-body decay, including the Tensor~\cite{Zemach:1965ycj} and helicity~\cite{Jacob:1959at} formalisms. 
In this analysis, the helicity formalism is followed to deal with \texttt{2F2P} decays, namely  $P_0\rightarrow P_1P_2P_3$ decays involving two spin-$1/2$ fermions and two (pseudo-)scalars.
According to the isobar model, the three-body decay is viewed as a two-step decay consisting of a weak decay followed by a strong decay for the problems interested in this analysis.
There are a total of $12$ degrees of freedom to describe the final-state kinematics, corresponding to the three four-momenta of particles $P_1$, $P_2$, and $P_3$.
After considering the constraints from energy-momentum conservation between the initial and final state, and that the final-state particles are on mass-shell, only five kinematic variables are independent. 
Three of the five remaining variables define the direction of the normal of the decay plane formed in the rest-frame of $P_0$ and the simultaneous rotation of final-state particles in the decay plane about the normal.
If $P_0$ is unpolarized (or specifically spinless), the distribution of these three variables carries no physical information. The other two variables define how the energy of the decaying particle is shared among the final state. They can be expressed by a two-body invariant mass $m_{ij}\equiv m(P_i,P_j)$, $P_i,P_j\in\{P_0,P_1,P_2\}, i\neq j$, and the helicity angle of the $P_iP_j$ system~\cite{Jacob:1959at}. 
The $m_{ij}$ distribution in the amplitude is usually described model dependently using the Breit-Wigner function~\cite{PDG} for a resonant contribution and an empirical smooth function for a non-resonant component. 
The helicity angle distribution is determined by Wigner functions~\cite{Jacob:1959at}, which depend on the spin  of the $P_iP_j$ system and rotation angles. 

Resonances may contribute to the amplitude in any two-body systems as $P_0\rightarrow R_{ij}P_k, R_{ij}\to P_iP_j$, such that the three-body decay is factorised into a sum of several chained two-body decays. In total there are three possible decay chains for a \texttt{2F2P} decay, $P^0\rightarrow R(P_1P_2)P_3$, $P^0\rightarrow R(P_1P_3)P_2$ and $P^0\rightarrow R(P_2P_3)P_1$. Resonance $R$ can be a fermion or a boson depending on the decay chain. In the following, the first and second chains are restricted to $R$ being fermions, and the third one is for $R$ being bosons.
Each decay in a chain is associated with a helicity-dependent complex coupling to describe the strength, named as the helicity coupling $H_{\lambda_a,\lambda_b}^P$ for a particle $P$ decaying into $a$ and $b$ with the helicity $\lambda_a$ and $\lambda_b$ respectively.  
The strong decay $R_{ij}\to P_iP_j$ conserves the parity symmetry, which requires couplings with positive helicities, and those with negative helicities are only different by a sign, $H^R_{\lambda_i,\lambda_j}=\eta_i\eta_j\eta_R(-1)^{J_R-J_i-J_j}H^R_{-\lambda_i,-\lambda_j}\equiv \eta^R_{ij} H^R_{-\lambda_i,-\lambda_j}$. 
Here  $J$ and $\eta$ are the corresponding  spin and parity quantum numbers of involved particles~\cite{Jacob:1959at}. 
For $J_0=J_i=1/2$, $J_j=J_k=0$, the two allowed strong couplings $H^R_{+1/2}=\eta^R H^R_{-1/2}$ can be absorbed into the weak helicity coupling of the $P_0\rightarrow R_{ij}P_k$ decay, $H^0_{+1/2}$ and $H^0_{-1/2}$, except for the $\eta^R=\pm1$ sign. Here, the other subscript of the helicity couplings, 0, is dropped. 
The sign $\eta^R$ measures the parity of the $R$ resonance.
For $J_i=J_j=1/2$, $J_0=J_k=0$, the four strong couplings are related to each other as $H^R_{+1/2,+1/2}=\eta^R H^R_{-1/2,-1/2}$, $H^R_{+1/2,-1/2}=\eta^R H^R_{-1/2,+1/2}$, while the single coupling of the $P_0\rightarrow R_{ij}P_k$ weak decay can be dropped. 
In the following, $H^R_+\equiv H^R_{+1/2,+1/2}$ and $H^R_-\equiv H^R_{-1/2,+1/2}$ are denoted.
For $J_0=J_k=1/2$, $J_i=J_j=0$, the four weak couplings are $H^0_{0,+1/2}, H^0_{-1,+1/2},H^0_{0,-1/2},H^0_{+1,-1/2}$, while the single coupling of the $R_{ij}\to P_iP_j$ strong decay can be dropped. These four couplings can be grouped into two parts, denoted as $H^0_{\pm}\equiv H^R_{0,\pm1/2}$ and $H^{'0}_\pm\equiv H^0_{\pm1,\mp1/2}$ respectively in the following.

As an example, the modulus squared of the unpolarized $\Lambda_b^0\rightarrow D^0p h^-$ decay amplitude, contributed by $R\to D^0p$ resonances, can be expressed as
\begin{equation}
\begin{aligned}
\text{PDF}=&\sum_{\lambda_{\Lambda_b}\lambda_p}\left|\sum_R H_{\lambda_{\Lambda_b}}^{\Lambda_b\to Rh}H^{R\to D^0p}_{\lambda_p}d^{J^R}_{\lambda_{\Lambda_b},\lambda_p}(\theta_p)F^R(m_{D^0 p})\right|^2\\
=&\left|\sum_R H^R_{+} d_{+1/2,+1/2}^{J^R}\left(\theta_p\right) F^R\left(m_{D^0 p}\right)\right|^2+\left|\sum_R \eta^R H^R_{-} d_{-1/2,-1/2}^{J^R}\left(\theta_p\right) F^R\left(m_{D^0 p}\right)\right|^2 \\
&+\left|\sum_R \eta^R H^R_{+} d_{+1/2,-1/2}^{J^R}\left(\theta_p\right)     F^R\left(m_{D^0 p}\right)\right|^2+\left|\sum_R H^R_{-} d_{-1/2,+1/2}^{J^R}\left(\theta_p\right) F^R\left(m_{D^0 p}\right)\right|^2,
\end{aligned}
\label{eqn:2F2P1}
\end{equation}
where $F^R(m_{D^0p})$ (shortened as $F^R$ in the following) denotes the model describing the mass distribution (the propagator) of resonance $R$ and $d^{J^R}_{\lambda_{\Lambda_b},\lambda_p}(\theta_p)$ is the Wigner small $d$-function depending on, the resonance spin $J^R$ the $\Lambda_b$ helicity $\lambda_{\Lambda_b}$, the proton helicity $\lambda_p$ and the helicity angle $\theta_p$ defined by the proton polar angle in $R$ rest-frame. 
For the second equality, the helicity coupling $H^{R\to D^0p}_{\lambda_p}$ is absorbed into $H_{\lambda_{\Lambda_b}}^{\Lambda_b\to Rh}$ leaving only the $\eta^R$ sign. 
Similarly, for the $B\rightarrow p \overline{p}h$ decay, considering only $R\to \overline{p}h$ resonances, the PDF is:
\begin{equation}
\begin{aligned}
\text{PDF} &=\sum_{\lambda_p \lambda_{\overline{p}}} \left| \sum_{R}H^{B\to Rp}_{ \lambda_p} H^{R\to \overline{p}h}_{\lambda_{\overline{p}}} d_{\lambda_p \lambda_{\overline{p}}}^{ J^R}\left(\theta_{\overline{p}}\right) F^R\right|^2,
\end{aligned}
\label{eqn:2F2P2}
\end{equation}
which has a form identical to the unpolarized $\Lambda_b^0\rightarrow D^0p h^-$ decay, by just replacing $\lambda_{\overline{p}}$ by $\lambda_{\Lambda_b}$. In addition, the unpolarized $\Lambda_b^0\rightarrow R(D^0 h^-)p$ decay has a PDF similar to that of the $B\rightarrow R(p \overline{p})h$ decay.

Helicity couplings of a PDF are unknown parameters to be determined by fitting the PDF to data.  It is apparent that the helicity couplings in a PDF have a non measurable global phase. In addition, the magnitude of one helicity coupling is not measurable due to the PDF normalization. These ambiguities can be removed by fixing $H_+=1+0i$  for the contribution of a reference resonance $R$.
Besides, for decays with all resonances in the same chain, the second equation of Eqn.~\ref{eqn:2F2P1} and the property of $d$-functions imply that, if $\{H_+^R, H_-^R\}$ is a solution, the simultaneous replacement $H_+^R\to\eta^R H_-^R, H_-^R\to\eta^RH_+^R$ for all resonances is also a solution. 
These two solutions can be distinguished by requiring, {\em e.g.}, $|H_-|<|H_+|$ for the reference $R$. Flipping the parity of all resonances gives another solution, which can be resolved using the parities of known particles. In section~\ref{section:MultipleSolution}, it will be shown that additional multiple solutions exist, and more requirements have to be imposed to obtain a unique solution.

\subsection{Multiple chains and the alignment angle}
In Eqns.~\ref{eqn:2F2P1} and~\ref{eqn:2F2P2},  resonances are only present in one decay chain, such that initial-state and final-state helicities for different sub-amplitudes are defined in the same reference frames, and are directly summed over incoherently.
For decays that have resonances in multiple decay chains, which is usually the case for multibody heavy-flavor decays, the final-state helicities in different chains are, however, not defined in the same reference frames. 
Additional rotations are needed to align the final-state helicity states, as has been discussed in Refs.~\cite{Chen:2017gtx,JPAC:2019ufm,Marangotto:2019ucc,Wang:2020giv}. In this manuscript, a new strategy is proposed to determine the alignment angles, based on the idea that the eigenstates of a particle spin are uniquely fixed by a coordinate system up to a common phase. 
Let $C^\text{ref}$ denote the coordinate system defining the helicity states in the reference chain and $C^\text{alt}$ denote that for another chain, which can be reached from $C^\text{ref}$ by the Euler rotation $C^\text{alt}=R(\alpha,\beta,\gamma)C^\text{ref}$. 
Then the state defined in $C^\text{alt}$, $|J,\lambda\rangle^\text{alt}$, is the linear superimposition of those in $C^\text{ref}$ as
\begin{equation}
    |J,\lambda\rangle^\text{alt} = \sum_{\lambda'}D_{\lambda'\lambda}^J(\alpha,\beta,\gamma)|J,\lambda'\rangle^\text{ref},\label{eq:spinIn2chains}
\end{equation}
where  $D_{\lambda'\lambda}^J(\alpha,\beta,\gamma)$ is the Wigner big $D$-function.
Specifically, for \texttt{2F2P} decays, when there are two decay chains of fermionic resonances, 
Eqn.~\ref{eqn:2F2P1} is then extended to

\begin{equation}
\begin{aligned}
|\mathcal{M}|^2=&\sum_{\lambda_{\Lambda_b}\lambda_p}\left|\{\mathcal{A}_{\lambda_{\Lambda_b}\lambda_p}\text{ of ref. chain}\} + \left(\langle 1/2,\lambda_p|\right)^\text{ref}
\left(\sum_{\lambda_p'}|1/2,\lambda_p'\rangle^\text{alt}\{ \mathcal{A}_{\lambda_{\Lambda_b}\lambda_p'}\text{ of alt. chain}\}\right)\right|^2\\
=&\sum_{\lambda_{\Lambda_b}\lambda_p}\left|\{\mathcal{A}_{\lambda_{\Lambda_b}\lambda_p}\text{ of ref. chain}\} + \sum_{\lambda_p'}D^{1/2}_{\lambda_p,\lambda_p'}(\alpha,\beta,\gamma)\{\mathcal{A}_{\lambda_{\Lambda_b}\lambda_p'}\text{ of alt. chain}\}
\right|^2,
\end{aligned}
\label{eqn:2F2P2Chains}
\end{equation}

where the completeness of  states $\sum_{\lambda_p'}|1/2,\lambda_p'\rangle\langle 1/2,\lambda_p'|=1$ has been inserted, and Eqn.~\ref{eq:spinIn2chains} has been used to obtain the second equation.  The Wigner $D$-functions align the proton helicities in the alternative decay chain to the reference chain.
The two coordinate systems $C^\text{ref}$ and  $C^\text{alt}$ have to be determined in the same reference frame as stressed in Ref.~\cite{LHCb:2015yax}. The Euler angles describing the rotations from one coordinate system to another one are calculated by the equations in appendix~\ref{app:EulerAngles}.

With helicity formalism, the coordinate systems for final-state particles are determined sequentially. Let's first define two arbitrary decay chains: the  sequence of decays $P_0\rightarrow R_{1}P_1, R_{1}\to R_2P_2, \cdots, R_n\to P_{n1}P_{n2}$ is referred to as the $R$ chain, and the $R ^{\prime}$ chain refers to  chained decays of $P_0\rightarrow R^{\prime}_{1}P^{\prime}_1, R^{\prime}_{1}\to R^{\prime}_2P^{\prime}_2, \cdots, R^{\prime}_n\to P_{n1}P^{\prime}_{n2}$. Here $P_{n1}$ and $P_{n2}^{(')}$ are final-state particles, while other $P_i$ can be final-state or intermediate particles. 
The alignment of $P_{n1}$ helicity states in the two chains is interesting here and those for other final-state particles can be determined following the same strategy. 
While the $R$ chain and the $R ^{\prime}$ chain involve different structures and compositions of intermediate states, the set of final states is identical. Given the coordinate system, $C^i$, for particle $R_i$ in the $R$ chain, the coordinate system for $R_{i+1}$, $C^{i+1}$, is obtained as the following~\cite{LHCb:2015yax}
\begin{equation}
\begin{aligned}
       \hat{z}^R_{i+1} &= I(\vec{p}^{R_{i+1},C^i}) \\
       \hat{y}^R_{i+1} &= I(\vec{z}^{C^i}\times \vec{p}^{R_{i+1},C^i})\\
       \hat{x}^R_{i+1} &= \hat{y}^R_{i+1}\times\hat{z}^R_{i+1},
\end{aligned}\label{eq:coord_all}
\end{equation}
where $\vec{p}^{R_{i+1},C^i}$ is the $R_{i+1}$ momentum in the rest-frame of $R_i$, and $I(\vec{v})$ takes the unit vector along $\vec{v}$. It is clear that within our convention the coordinate systems for the particle $R_i$ and $P_i$ are anti-parallel in $z$ and $y$ directions, and are parallel in the $x$ direction. For this convention, particles $R_i$ and $P_i$ are placed symmetrically in the amplitude, and the parity symmetry relation for helicity couplings is preserved. Given an initial coordinate system for $P_0$, that for any  particle $i$ in the $R$ ($R'$) chain, $C^{i(R)}$ ($C^{i(R')}$),  can be uniquely defined by repeating Eqn.~\ref{eq:coord_all}. 
It is apparent that a particle's coordinate system is chain dependent.

The systems $C^{i(R')}$ and $C^{i(R)}$ are defined in two different reference frames  since they are reached from the $P_0$ rest-frame by different paths of boosts. 
They must be brought to the same reference frame before helicity alignment angles can be calculated from the two systems. The list of sequential boosts that bring the $P_0$ rest-frame to the $P_{f}$ rest-frame in the $R$ chain is defined as $\{B_1^R, B_2^R\cdots B_{f}^R\}$ and that in the $R'$ chain is defined as $\{B_1^{R'}, B_2^{R'}\cdots B_{f}^{R'}\}$, where $f\equiv n1$ for a special case of the previous section. For example, for the $P_0\rightarrow R_{12}P_3, R_{12}\to P_1P_2$ decay, a boost $B^{R_{12}}_1$ from the $P_0$ rest-frame to the $R$ rest-frame followed by a boost $B^{R_{12}}_2$ from the $R_{12}$ rest-frame to the $P_1$ rest-frame reaches the $P_1$ reference frame in the $R_{12}$ chain.
Now the three coordinates of $C^{f(R')}$  are extended to four vectors by adding an arbitrary time component (0 for example), 
$x^{f(R')}\equiv(\hat{x}^{f(R')},0)$, 
$y^{f(R')}\equiv(\hat{y}^{f(R')},0)$ and 
$z^{f(R')}\equiv(\hat{z}^{f(R')},0)$. 
The vector $x^{f(R')}$ can be transformed to the particle $P_f$ rest-frame through the  $R$ chain as
\begin{equation}
x^{f(R'\to R)} 
=\left(B_f^R\cdots B_2^RB_1^R\right)\left(B_f^{R'}\cdots B_2^{R'}B_1^{R'}\right)^{-1}x^{f(R')},
\end{equation}
similarly for $y^{f(R'\to R)}$ and $z^{f(R'\to R)}$. Here a boost $B$ is in the representation for Lorentz vectors and is only determined by the momentum of a particle whose rest frame is to be reached. Taking the space components of $x^{f(R'\to R)},y^{f(R'\to R)}$ and $z^{f(R'\to R)}$, the coordinate system for $P_f$ in $R'$ chain transformed to the $R$ chain, $C^{f(R'\to R)}$,  is obtained. The Euler rotations that bring $C^{f(R)}$ to $C^{f(R'\to R)}$  give the alignment angles needed in Eqn.~\ref{eqn:2F2P2Chains}, where the $R$ chain is the reference. The same procedure is repeated to determine the helicity alignment angles of all final-state particles.

In the case of a \texttt{2F2P} decay, the decay chain can be simplified to $P_0\rightarrow R_{12}P_3, R_{12}\to P_1P_2$. If the decaying particle has $J=1/2$, {\em e.g.} the $\Lambda_b^0$ baryon in Eqn.~\ref{eqn:2F2P1}, alignments of its helicities in different chains are also needed. It is properly considered by choosing the same initial coordinate system for $P_0$ for all decay chains. Similarly for the $B\rightarrow p\overline{p}h$ decay in Eqn.~\ref{eqn:2F2P2},  helicity alignments should be applied for both $p$ and $\overline{p}$.  Thus for \texttt{2F2P} decays, two additional Wigner-$D$ rotations are needed for the alternative chain in Eqn.~\ref{eqn:2F2P2Chains}.

\begin{figure}
\centering
\resizebox{0.4\textwidth}{!}{%
  \includegraphics{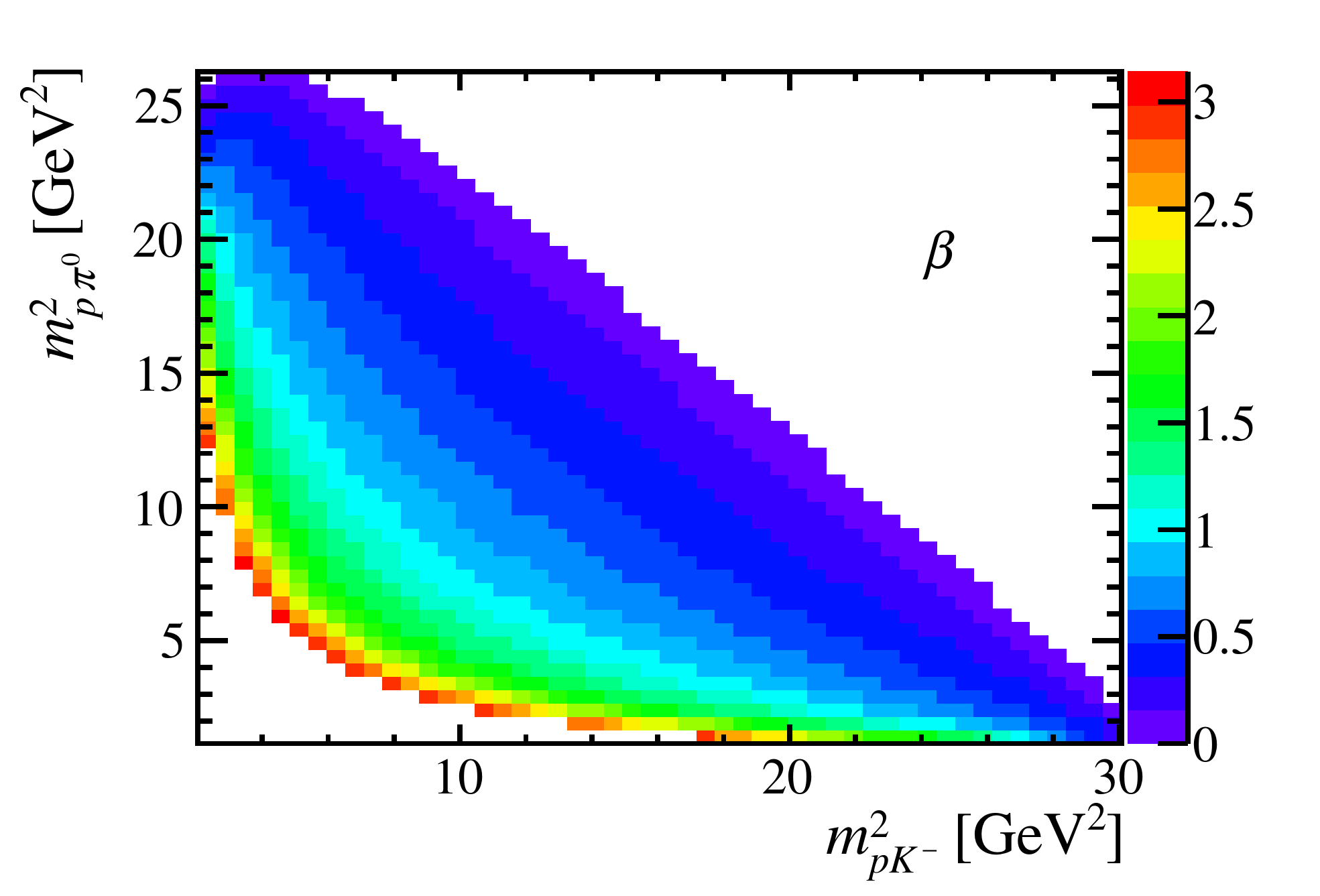}
}
\caption{The distribution of the Euler angle $\beta\in [0,\pi]$ that aligns the proton helicity in two decay chains, as a function of the phase-space position.}
\label{fig:BetaLb2pKpi0}       
\end{figure}

Before concluding this section, our approach is compared with other methods, {\em e.g.} that in Ref.~\cite{JPAC:2019ufm}, for unpolarized $\Lambda_b^0\to p K^- \pi^0$ decays. For this decay, three chains are possible to reach the final state, and alignments of the proton helicity states in different chains are needed. Numeric values show that  a $\beta$ rotation about the $y$-axis (defined to be the normal of the decay plane in the $\Lambda_b^0$ rest-frame) is identical in both methods to align two chains. However, in our method, a rotation by $\pi$ around the $z$-axis may be needed since the  $y$-axis of the proton coordinate system may flip sign in different chains and the $\beta$ angle is restricted to the range $[0,\pi]$. In Fig.~\ref{fig:BetaLb2pKpi0}, the distribution of the $\beta$ angle that aligns the proton helicity in the $\Lambda_b^0\to R(p K^-) \pi^0$ chain (reference) to that in the $\Lambda_b^0\to R(p  \pi^0) K^-$ chain as a function of the two-body invariant-mass squared, $m^2_{p K^-}$ and $m^2_{p \pi^0}$, is shown. At low $m_{K^-\pi^0}$, the two proton helicities are almost parallel such that a small $\beta$ rotation is needed, but they are almost anti-parallel  at high $m_{K^-\pi^0}$, demanding a large rotation $\beta$ about the $y$-axis.

\section{Multiple solutions of amplitude fits}
\label{section:MultipleSolution}
In this section the $\Lambda_b^0\to p h^-h^0$ decay is taken as an example to demonstrate why there are multiple solutions in amplitude fits of \texttt{2F2P} decays and how to resolve it. 
As a start point, the scenario where only intermediate states decaying into $ph^-$ final state are allowed and only two resonances, with  $J^\eta=1/2^-$, contributing to the amplitude are considered. The study is later extended to decays with more states and arbitrary $J^\eta$. Finally, decays with more than one chain will be discussed.
\subsection{One chain with two $J^\eta=1/2^-$ resonances}

For only one decay chain with $ph^-$ resonances, the PDF of the $\Lambda_b^0\to p h^-h^0$ decay can be obtained using Eqn.~\ref{eqn:2F2P1} as:

\begin{equation}
\begin{aligned}
\text{PDF}
=&\left|\sum_R H^R_{+} d_{+1/2,+1/2}^{1/2}\left(\theta_p\right) F^R\right|^2
+\left|\sum_R  H^R_{-} d_{+1/2,+1/2}^{1/2}\left(\theta_p\right) F^R\right|^2 \\
&+\left|\sum_R H^R_{+} d_{-1/2,+1/2}^{1/2}\left(\theta_p\right)     F^R\right|^2+\left|\sum_R H^R_{-} d_{-1/2,+1/2}^{1/2}\left(\theta_p\right) F^R\right|^2,
\end{aligned}
\label{eqn:ppbaronechannel}
\end{equation}
where $F^R$ is the $ph^{-}$ invariant mass distribution. The relations for Wigner-$d$ functions 
    $d^j_{m,m'}=d^j_{-m',-m}=(-1)^{m-m^\prime}d_{m',m}^j$
and the parity coefficient $\eta^R=1$ for $J^\eta=1/2^-$ resonances in this special case have been used. It is noticed that since the real $d$-functions are identical for all resonances, they can be pulled outside of the modulus, giving
    \begin{equation}
\begin{aligned}
\text{PDF} 
=&\left(\left|\sum_R H^R_{+}  F^R\right|^2
+\left|\sum_R  H^R_{-}  F^R\right|^2\right)\left[d_{+1/2,+1/2}^{1/2}\left(\theta_p\right)\right]^2 
+\left(\left|\sum_R H^R_{+}      F^R\right|^2+\left|\sum_R H^R_{-} F^R\right|^2\right)\left[d_{-1/2,+1/2}^{1/2}\left(\theta_p\right)\right]^2\\
=&\left(\left|\sum_R H^R_{+}  F^R\right|^2
+\left|\sum_R  H^R_{-}  F^R\right|^2\right)\left\{\left[d_{+1/2,+1/2}^{1/2}\left(\theta_p\right)\right]^2+
\left[d_{-1/2,+1/2}^{1/2}\left(\theta_p\right)\right]^2
\right\}\\
=&
\left|\sum_R (A^R+iB^R)  F^R\right|^2
+\left|\sum_R  (C^R+iD^R)  F^R\right|^2,
\end{aligned}
\label{eqn:ppbaronechannel1}
\end{equation}

where $A^R\equiv\mathcal{R}e(\mathcal{H}_+^R)$, $B^R\equiv\mathcal{I}m(\mathcal{H}^R_+)$, $C^R\equiv\mathcal{R}e(\mathcal{H}^R_-)$, $D^R\equiv\mathcal{I}m(\mathcal{H}^R_-)$ represent the respective real and imaginary parts of the complex couplings $H_+$ and $H_-$ \footnote{For the discussions below, the $\{A, B, C, D\}$ notations are not mandatory, but they help to make the problem easier to understand.}.
For the last equation, the explicit expressions of Wigner-$d$ functions are applied as summarised in Tab.~\ref{table:wignerdexpand} for a few examples.

\begin{table}
\caption{Examples of Wigner-$d$ functions.}
\label{table:wignerdexpand}       
\centering
\begin{tabular}{l|c|c}
\hline\noalign{\smallskip}
$J^R$ & $d_{+1 / 2,+1 / 2}^{J^R}\left(\theta\right)$ & $d_{-1 / 2,+1 / 2}^{J^R}\left(\theta\right)$ \\
\noalign{\smallskip}\hline\noalign{\smallskip}
$1 / 2$ & $\cos \frac{\theta}{2}$ & $-\sin \frac{\theta}{2}$ \\
$3 / 2$ & $\frac{1}{2} \cos \frac{\theta}{2}\left(3 \cos \theta-1\right)$ & $-\frac{1}{2} \sin \frac{\theta}{2}\left(3 \cos \theta+1\right)$ \\
$5 / 2$ & $\frac{1}{2} \cos \frac{\theta}{2}\left(5 \cos ^2 \theta-2 \cos \theta_p-1\right)$ & $-\frac{1}{2} \sin \frac{\theta}{2}\left(5 \cos ^2 \theta+2 \cos \theta-1\right)$\\
\noalign{\smallskip}\hline
\end{tabular}
\end{table}

The invariant mass distributions $F^R(m_{ph^{-}})$ is usually taken to be the form of the Breit-Wigner distribution,
\begin{equation}
    F^R(m_{ph^{-}}\mid \mu_R,g_R)=\dfrac{1}{\mu_R^2-m_{ph^{-}}^2-i\mu_R\Gamma_R(m_{ph^{-}}\mid \mu_R,g_R)},
\end{equation}
where $\mu_R$ and $g_R$ are the mass and natural width of the resonance $R$. Then Eqn.\ref{eqn:ppbaronechannel1} is transformed to
\begin{equation}
\begin{aligned}
\text{PDF}
=&\left|\sum_R \frac{(A^R+iB^R)}{F^R_\mathcal{R}-iF^R_\mathcal{I}}  \right|^2+\left|\sum_R \frac{(C^R+iD^R )}{F^R_\mathcal{R}-iF^R_\mathcal{I}} \right|^2,
\end{aligned}
\label{eqn:ppbaronchannelexpand}
\end{equation}
where $F^R_\mathcal{R}\equiv \mu_R^2-m_{ph}^2$, $F^R_\mathcal{I}\equiv\mu_R\Gamma_R(m_{ph^{-}}\mid \mu_R,g_R)$ take the real and imaginary parts of the denominator of the Breit-Wigner distribution. It is noted that  Eqn.~\ref{eqn:ppbaronchannelexpand} holds for any number of $ph^-$ resonances with $J^\eta=1/2^{-}$.

For two resonances, by multiplying $G_1^2\times G_2^2\equiv\left|F^{R_1}_\mathcal{R}-iF^{R_1}_\mathcal{I}\right|^2\times\left|F^{R_2}_\mathcal{R}-iF^{R_2}_\mathcal{I}\right|^2$ on both sides of Eqn.~\ref{eqn:ppbaronchannelexpand}results in

    \begin{equation}
\begin{aligned}
\text{PDF}\times G_1^2 G_2^2=&\left|\left(A^{R_1}+iB^{R_1}\right)\left(F^{R_2}_\mathcal{R}-iF^{R_2}_\mathcal{I}\right)+\left(A^{R_2}+iB^{R_2}\right)\left(F^{R_1}_\mathcal{R}-iF^{R_1}_\mathcal{I}\right)\right|^2\\
&+\left|\left(C^{R_1}+iD^{R_1}\right)\left(F^{R_2}_\mathcal{R}-iF^{R_2}_\mathcal{I}\right)+\left(C^{R_2}+iD^{R_2}\right)\left(F^{R_1}_\mathcal{R}-iF^{R_1}_\mathcal{I}\right)\right|^2 \\
=&\left[(A^{R_1})^2+(B^{R_1})^2+(C^{R_1})^2+(D^{R_1})^2\right]\left[(F^{R_2}_\mathcal{R})^2+(F^{R_2}_\mathcal{I})^2\right]\\
&+\left[(A^{R_2})^2+(B^{R_2})^2+(C^{R_2})^2+(D^{R_2})^2\right]\left[(F^{R_1}_\mathcal{R})^2+(F^{R_1}_\mathcal{I})^2\right] \\
&+2\left(A^{R_1} A^{R_2}+B^{R_1} B^{R_2}+C^{R_1} C^{R_2}+D^{R_1} D^{R_2}\right)\left(F^{R_1}_\mathcal{R}F^{R_2}_\mathcal{R}+ F^{R_1}_\mathcal{I}F^{R_2}_\mathcal{I}\right) \\
&+2\left(A^{R_1} B^{R_2}-A^{R_2} B^{R_1}+C^{R_1} D^{R_2}-C^{R_2} D^{R_1}\right)\left(
F^{R_2}_\mathcal{R}F^{R_1}_\mathcal{I}- F^{R_1}_\mathcal{R}F^{R_2}_\mathcal{I}\right).
\end{aligned}
\label{eqn:ppbaronchannelexpandtimes}
\end{equation}

The helicity coupling parameters $\{A^R, B^R, C^R, D^R\}$ are unknowns, to be determined from data.
It is noticed that only combinations of these coupling parameters appear in the PDF. For simplicity,  the following notations are made:

\begin{equation}
\begin{aligned}
    P_{1\cdot 1}\equiv&\, (A^{R_1})^2+(B^{R_1})^2+(C^{R_1})^2+(D^{R_1})^2=|H_+^{R_1}|^2+|H_-^{R_1}|^2,\\
    P_{2\cdot 2}\equiv&\, (A^{R_2})^2+(B^{R_2})^2+(C^{R_2})^2+(D^{R_2})^2=|H_+^{R_2}|^2+|H_-^{R_2}|^2,\\
    P_{1\cdot 2}\equiv&\, A^{R_1}A^{R_2}+B^{R_1}B^{R_2}+C^{R_1}C^{R_2}+D^{R_1}D^{R_2}=\mathcal{R}e(H_+^{R_1*}H_+^{R_2}+H_-^{R_1*}H_-^{R_2}),\\
    P_{1\times 2}\equiv&\, A^{R_1}B^{R_2}-A^{R_2}B^{R_1}+C^{R_1}D^{R_2}-C^{R_2}D^{R_1}=\mathcal{I}m(H_+^{R_1*}H_+^{R_2}+H_-^{R_1*}H_-^{R_2}),
\end{aligned}
\label{eqn:combination}
\end{equation}

which can be followed to define $P_{m\cdot n}$ and $ P_{m\times n}$ for any two resonances. These four $P$ parameters can be determined simultaneously by fitting Eqn.~\ref{eqn:ppbaronchannelexpandtimes} to data, however, the set of $\{A^R, B^R, C^R, D^R\}$ parameters are redundant and can't all be determined individually. This is the multiple solution problems particularly interesting in this manuscript.

The multiple-solution problem for \texttt{2F2P} decays is rephrased as, given an arbitrary set of $\{A^R, B^R, C^R, D^R\}$ parameters, other sets result in the same $P$ parameters. For two resonances of the same $J^\eta$ in a single decay chain, according to Eqn.~\ref{eqn:combination}, not all the free helicity couplings can be determined, four of which must be fixed to reach a stable fit result. For a maximum likelihood fit, the requirement to normalize the PDF removes one additional parameter. This trivial multiple-solution problem is always resolved by fixing  $\left|H_+^R\right|=1$ for the reference resonance $R$ and is not considered anymore in the remaining part of this manuscript. For the special case that only one resonance is present in the decay, only $P_{1\cdot1}$ remains in Eqn.~\ref{eqn:combination}, and all four parameters can be fixed after the normalization.
Later a strategy to systematically fix redundant helicity couplings to remove multiple solutions will be propsed.

\subsection{One decay chain with $N$ arbitrary resonances}
\label{section:onechannel}
Eqn.~\ref{eqn:ppbaronchannelexpandtimes} can be extended to an arbitrary number of $N$ resonances with varied $J^\eta$ in the same decay chain as 

    \begin{equation}
\begin{aligned}
\text{PDF}\times\prod_{R} G_R^2=&\left|\sum_R (A^R+iB^R) d_{+1 / 2,+1 / 2}^{J^R}\left(\theta\right)\prod_{R',R'\neq R} \left(F^{R'}_\mathcal{R}-iF^{R'}_\mathcal{I}\right)  \right|^2\\
&+\left|\sum_R \eta^R(A^R+iB^R) d_{-1 / 2,+1 / 2}^{J^R}\left(\theta\right)\prod_{R',R'\neq R} \left(F^{R'}_\mathcal{R}-iF^{R'}_\mathcal{I}\right)  \right|^2\\
&+\left|\sum_R \eta^R(C^R+iD^R)d_{+1 / 2,+1 / 2}^{J^R}\left(\theta\right)\prod_{R',R'\neq R} \left(F^{R'}_\mathcal{R}-iF^{R'}_\mathcal{I}\right) \right|^2\\
&+\left|\sum_R (C^R+iD^R)d_{-1 / 2,+1 / 2}^{J^R}\left(\theta\right)\prod_{R',R'\neq R} \left(F^{R'}_\mathcal{R}-iF^{R'}_\mathcal{I}\right)  \right|^2
\end{aligned}
\label{eqn:ppbaronchannelexpandtimesN}
\end{equation}
The list of $P$ parameters appears in the PDF can be defined using those $\{A^R,B^R,C^R,D^R\}$ in Eqn.~\ref{eqn:ppbaronchannelexpandtimesN} as
    \begin{equation}
\begin{aligned}
    P_{m\cdot n}=&\, \eta^{R_m}\eta^{R_n}A^{R_m}A^{R_n}+\eta^{R_m}\eta^{R_n}B^{R_m}B^{R_n}+C^{R_m}C^{R_n}+D^{R_m}D^{R_n},\\
    P_{m\times n}=&\, \eta^{R_m}\eta^{R_n}A^{R_m}B^{R_n}-\eta^{R_m}\eta^{R_n}A^{R_n}B^{R_m}+C^{R_m}D^{R_n}-C^{R_n}D^{R_m},
\end{aligned}
\label{eqn:combinationsN}
\end{equation}

where $R_m$ and $R_n$ represent the $m^\text{th}$ and $n^\text{th}$ resonances in the list. The parity coefficients $\eta^R$ are present in the PDF since they are not identical for all resonances. From Eqn.~\ref{eqn:combinationsN}, it is clear that the interchange $H_+^R\leftrightarrow \eta^RH_-^R$, or $A^R\leftrightarrow\eta^R C^R, B^R\leftrightarrow\eta^R D^R$, for all $R$ at once gives an equivalent PDF. Eqn.~\ref{eqn:combinationsN} is back to Eqn.~\ref{eqn:combination} without $\eta^R$ coefficients by redefining $\tilde{A}^R\equiv\eta^RA^R$ and $\tilde{B}^R\equiv\eta^RB^R$. 

For $N$ resonances in the same chain, a total of $N^2$  $P$ parameters can be defined from the $4N$ degrees of freedom in helicity couplings. 
Even though for $N\geq4$, the number of $P$ parameters is not less than the number of independent helicity parameters, still not all helicity parameters can be uniquely determined from data. Let's define a matrix $H$ of helicity parameters as
\begin{equation}
\begin{aligned}
H\equiv
&\left(\begin{array}{llll}
\tilde{A}^{R_1}+i\tilde{B}^{R_1} & \tilde{A}^{R_2}+i\tilde{B}^{R_2} &\cdots & \tilde{A}^{R_N}+i\tilde{B}^{R_N}\\
C^{R_1}+iD^{R_1} & C^{R_2}+iD^{R_2} &\cdots & C^{R_N}+iD^{R_N} 
\end{array}\right),
\end{aligned}
\label{eqn:H}
\end{equation}
then $H^\dagger H$ would give all the $P_{m\cdot n}$ and $P_{m\times n}$ terms, as $(H^\dagger H)_{m,n}=P_{m\cdot n}-iP_{m\times n}$.

For any non-trivial $2\times2$ unitary matrix $U\neq I$, $U^\dagger U=I$, the identity $(U\!H)^\dagger U\!H=H^\dagger U^\dagger U\!H=H^\dagger H$ holds, such that if the $H$ matrix elements provide a solution for a fit, the elements of $U\!H$ provide another solution. The  arbitrariness of $U$ leads to the multiple-solution problem and in general any two solutions are linked by a unitary transformation. The $2\times2$ matrix $U$ belongs to the $U(2)$ group, which has four independent parameters. The $U$ matrix can be properly chosen  to eliminate four free parameters in the $H$ matrix to reach a definite solution. 

For example, given a solution in Eqn.~\ref{eqn:H}, a matrix $U$ in the specific form can be calculated as
\begin{equation}
U=\left(\begin{array}{cc}
1 & 0\\
0 & \exp^{i\psi}
\end{array}\right)\times\frac{1}{\mathcal{N}}\left(\begin{array}{cc}
\tilde{A}^{R_1}-i\tilde{B}^{R_1} & C^{R_1}-iD^{R_1} \\
-C^{R_1}-iD^{R_1} & \tilde{A}^{R_1}+i\tilde{B}^{R_1} 
\end{array}\right),
\label{eqn:rotation}
\end{equation}
where $\mathcal{N}=\sqrt{(\tilde{A}^{R_1})^2+(\tilde{B}^{R_1})^2+(C^{R_1})^2+(D^{R_1})^2}$ and $\psi$ is a phase factor to be determined later.  It is easily verified that $U$ is unitary. Then,

    \begin{equation}
H^{''}\equiv U\!H
=\frac{1}{\mathcal{N}}\left(\begin{array}{lll}
\mathcal{N}^2 & (\tilde{A}^{R_1}-i\tilde{B}^{R_1})(\tilde{A}^{R_2}+i\tilde{B}^{R_2})+(C^{R_1}-iD^{R_1})(C^{R_2}+iD^{R_2}) &\cdots  \\
0 & \exp^{i\psi}\left[(\tilde{A}^{R_1}+i\tilde{B}^{R_1})(C^{R_2}+iD^{R_2})-(\tilde{A}^{R_2}+i\tilde{B}^{R_2})(C^{R_1}+iD^{R_1})\right] &\cdots
\end{array}\right),
\label{eqn:RH}
\end{equation}

where $B^{''R_1}=C^{''R_1}=D^{''R_1}=0$ have been obtained and $\psi$ can be tuned to set $D^{''R_2}=0$, {\em i.e.} four redundant parameters have been fixed. Of course, one can choose another $U$ matrix to get the preferred form of $H^{''}$. Therefore, it is proved that it is always possible to fix four parameters to zero  if  resonances are limited to only one chain, reducing the total unknown helicity coupling parameters to $4N-4$. The detailed form of the transformed $H$ matrix is provided in appendix~\ref{eqn:PSingleChain}.

After the fix of $4$ helicity coupling parameters, the number of independent $P$ terms is always no less than the number of remaining free parameters, $N^2-(4N-4)=(N-2)^2\geq0$. It is noted that $P_{m\times n}$ terms only appear in the PDF if the mass distribution of either the $m$ or $n$  component of the amplitude is complex ({\em e.g.} the Breit-Wigner function). If all $P_{m\times n}$ terms disappear, then the number of $P$ terms reduces to $N+N(N-1)/2=N(N+1)/2$, which is smaller than the number of free coupling parameters $4N-4$ for $N<6$, suggesting additional parameters can be fixed.  For example, for two resonances, the following $P$ terms are obtained without $P_{1\times 2}$
\begin{equation}
\begin{aligned}
    P_{1\cdot 1}\equiv&\, (\tilde{A}^{R_1})^2,\\
    P_{2\cdot 2}\equiv&\, (\tilde{A}^{R_2})^2+(\tilde{B}^{R_2})^2+(C^{R_2})^2,\\
    P_{1\cdot 2}\equiv&\, \tilde{A}^{R_1}\tilde{A}^{R_2},
\end{aligned}
\label{eqn:combination_}
\end{equation}
where other $4$ coupling parameters have been set to zero following Eqn.~\ref{eqn:RH}.
The parameter $\tilde{A}^{R_1}$ is fixed by $P_{1\cdot 1}$, $\tilde{A}^{R_2}$ is fixed by $P_{1\cdot 2}$ and $\tilde{A}^{R_1}$, while $P_{2\cdot 2}$ can only additionally determine $(\tilde{B}^{R_2})^2+(C^{R_2})^2$. One can fix $C^{R_2}=0$ to reach a definite solution. For three resonances in the same chain, one can set $C^{R_2}=C^{R_3}=0$, and so on. This specific decay is not discussed anymore in the following.

\subsection{Multiple decay chains}
\label{subsection:multi}
In section \ref{section:onechannel}, resonances are only considered in  the first decay chain (referred to as the reference channel in the following). In this case the $H_+$ ($H_-$) coupling for one resonance only couples to the $H_+$ ($H_-$) of other resonances but not the $H_-$ ($H_+$), which can be seen from Eqns.~\ref{eqn:2F2P1} and~\ref{eqn:2F2P2}, wherein the same modulus the helicity for the initial (final) state is the same for different resonances. However when more than one chain is included in the amplitude, as can be seen from Eqn.~\ref{eqn:2F2P2Chains}, the $H_+$ ($H_-$) couplings of resonances in the reference chain also couple to the $H_-$ ($H_+$) couplings of another chain due to the realignment of initial- and final-state helicities. One would naively think that all the relative phases are fixed as phases of $H_+$ and $H_-$ of the alternative chain are fixed to a $H_+$ in the reference chain through one of the moduli. Then the phases of $H_+$ and $H_-$ in the alternative chain fix the $H_-$ of the reference chain in other moduli. However, it will be shown that apart from a global phase, which can be used to fix $\mathcal{I}m(H^{R_1}_+)=0$, there is a second degree of freedom to set $\mathcal{I}m(H^{R_1}_-)=0$, where $R_1$ is an arbitrary reference resonance.

For a \texttt{2F2P} decay with resonances in two different chains,  two kinds of combinations of helicity coupling parameters are present in the PDF. Without the loss of generality, the discussions are based on the first two chains, where intermediate resonances are fermions.
Besides those $P$ terms in Eqn.~\ref{eqn:combination} for   resonances in the same chain,  the following new terms appear between resonances in two different chains
\begin{equation}
    \begin{aligned}
    P^S_{m\cdot n}\equiv&\, \mathcal{R}e(H_+^{R_m*}H_+^{R'_n}+\text{S}\times H_-^{R_m*}H_-^{R'_n}),\\
    P^S_{m\times n}\equiv&\, \mathcal{I}m(H_+^{R_m*}H_+^{R'_n}+\text{S}\times H_-^{R_m*}H_-^{R'_n}),\\
    Q^S_{m\cdot n}\equiv&\, \mathcal{R}e(H_+^{R_m*}H_-^{R'_n}+\text{S}\times  H_-^{R_m*}H_+^{R'_n}),\\
    Q^S_{m\times n}\equiv&\, \mathcal{I}m(H_+^{R_m*}H_-^{R'_n}+ \text{S}\times H_-^{R_m*}H_+^{R'_n}),\\
    \end{aligned}
    \label{eqn:PQformulti}
\end{equation}
where and $R$ and $R'$ belong to two different chains, $\text{S}=+1$ or $-1$ depending on $R_m, R_n$ parity parameters $\eta$ and if a $\pi$ angle is needed to align initial/final-state helicities in the two decay chains. The detailed derivation of terms in Eqn.~\ref{eqn:PQformulti} can be found in appendix~\ref{appendix:multi}. As is seen, the $Q$ terms represent interferences of positive and negative helicities of two separate amplitude components.

For $M$ resonances in the reference chain and $N$ resonances in the alternative chain, the total number of independent $P$ and $Q$ terms is $M^2+N^2+4MN$. To generate all these terms, two $H$ matrices similar to that in Eqn.~\ref{eqn:H} are needed, defined as

\begin{equation}
\begin{aligned}
H_P\equiv
&\left(\begin{array}{llllllll}
H^{R_1}_+ & H^{R_2}_+ &\cdots  & H^{R_M}_+ & H^{R'_{M+1}}_+& H^{R'_{M+2}}_+ &\cdots& H^{R'_{M+N}}_+\\
H^{R_1}_- & H^{R_2}_- &\cdots  & H^{R_M}_- & H^{R'_{M+1}}_-& H^{R'_{M+2}}_- &\cdots& H^{R'_{M+N}}_-
\end{array}\right),\\
H_Q\equiv
&\left(\begin{array}{llllllll}
H^{R_1}_+ & H^{R_2}_+ &\cdots  & H^{R_M}_+ & H^{R'_{M+1}}_-& H^{R'_{M+2}}_- &\cdots& H^{R'_{M+N}}_-\\
H^{R_1}_- & H^{R_2}_- &\cdots  & H^{R_M}_- & H^{R'_{M+1}}_+& H^{R'_{M+2}}_+ &\cdots& H^{R'_{M+N}}_+
\end{array}\right),
\end{aligned}
\label{eqn:H2Chain2}
\end{equation}

where the $\text{S}$  coefficient is omitted for simplicity as they won't
affect the conclusion. The first $M$ columns of $H_P$ and $HQ$ are identical, and for $M+1$ to $M+N$ columns, the first and second rows of $H_P$ and $H_Q$ are swapped.
It is verified that $(H_P^\dagger H_P)_{m,n}=P_{m\cdot n}-iP_{m\times n}$ and $(H_Q^\dagger H_Q)_{m\leq M,n>M}=Q_{m\cdot n}-iQ_{m\times n}$, giving all the desired $P$ and $Q$ terms.

The question is to find a unitary matrix $U$ with  $H_P'\equiv UH_P$ and $H_Q'\equiv UH_Q$, such that
\begin{equation}\begin{aligned}
    (H_P')_{1j}=&(H_Q')_{1j} \\
    (H_P')_{2j}=&(H_Q')_{2j} 
\end{aligned}\label{eqn:HPQRef}\end{equation}
for $j\leq M$ (the reference chain), and
\begin{equation}\begin{aligned}
    (H_P')_{1j}=&(H_Q')_{2j} \\
    (H_P')_{2j}=&(H_Q')_{1j} 
\end{aligned}\label{eqn:HPQAlt}\end{equation}
for $M<j\leq M+N$ (the alternative chain). Namely, the  elements of transformed $H_P$ and $H_Q$ matrices are related. Eqn.~\ref{eqn:HPQRef} holds automatically since $(H_P)_{ij}=(H_Q)_{ij}$ for $j\leq M$, corresponding to the reference chain. For the alternative chain, Eqn.~\ref{eqn:HPQAlt} can be translated to the requirements
\begin{equation}
\begin{aligned}
U
\left(\begin{array}{l}
X\\Y
\end{array}\right)
\end{aligned}
=
\begin{aligned}
\left(\begin{array}{l}
X'\\Y'
\end{array}\right),
\end{aligned}\,\,\,
\begin{aligned}
U
\left(\begin{array}{l}
Y\\X
\end{array}\right)
\end{aligned}
=
\begin{aligned}
\left(\begin{array}{l}
Y'\\X'
\end{array}\right),
\end{aligned}
\label{eqn:UXY}
\end{equation}
for any $X,Y$, which requires 
\begin{equation}
\left(\begin{array}{ll}
0 & 1\\
1 & 0
\end{array}\right)U
\left(\begin{array}{ll}
0 & 1\\
1 & 0
\end{array}\right)=U.
\label{eqn:UXY2}
\end{equation}
For a generic unitary matrix of four degrees of freedom,
\begin{equation}
U=\left(\begin{array}{ll}
\alpha & \beta\\
-\beta^*e^{i\gamma} & \alpha^*e^{i\gamma}
\end{array}\right),
\label{eqn:Umatrix}
\end{equation}
where $\alpha$ and $\beta$ are complex and $|\alpha|^2+|\beta|^2=1$, Eqn.~\ref{eqn:UXY2} implies
\begin{equation}
\begin{aligned}
\alpha=\alpha^*e^{i\gamma},\\
\beta=-\beta^*e^{i\gamma},\\
\end{aligned}
\label{eqn:UmatrixConstrained}
\end{equation}
which determine $\gamma=2\arg{\alpha}$ and  $\arg{\beta}=\arg{\alpha}\pm \pi/2$, eliminating two degrees of freedom. Now the reduced $U$ matrix has the generic form,
\begin{equation}
U(\phi,t)=e^{i\phi}\left(\begin{array}{ll}
\cos(t) &  i \sin(t)\\
 i\sin(t) &\cos(t)
\end{array}\right),
\label{eqn:UmatrixSpecial}
\end{equation}
with  $\phi\in(0,2\pi]$ and $t\in(-\pi/2,\pi/2]$. Next it will be demonstrated that $(\phi,t)$ parameters can be tuned to set $\mathcal{I}m(H_+^{R_1})=\mathcal{I}m(H_-^{R_1})=0$.

Multiplying the reduced $U$ matrix in Eqn.~\ref{eqn:UmatrixSpecial} to $H_P$ (only the first column is needed without the loss of generality) reaches
\begin{equation}
\begin{aligned}
U(\phi,t) H_P=&e^{i\phi}\left(\begin{array}{ll}
\cos(t) &  i \sin(t)\\
i\sin(t) &\cos(t)
\end{array}\right)\times
\left(\begin{array}{ll}
H_+ & \cdots\\
H_- & \cdots
\end{array}\right)\\
=&e^{i\phi}
\left(\begin{array}{ll}
\cos(t) H_+ +i \sin(t) H_- & \cdots\\
\cos(t) H_- +i \sin(t) H_+ & \cdots
\end{array}\right)\\
\equiv &e^{i\phi}
\left(\begin{array}{ll}
H_+' & \cdots\\
H_-'& \cdots
\end{array}\right),
\end{aligned}
\label{eqn:UHP}
\end{equation}
where the superscript $R_1$ has been omitted.
The parameter $t$ is adjusted to make $H_+$ and $H_-$  have a common phase $\delta$, which then can be removed by choosing $\phi=-\delta$. It requires

    \begin{eqnarray}
\tan(\delta)=\frac{
\mathcal{I}m(H_+)\cos(t)+ \mathcal{R}e(H_-)\sin(t)}
{\mathcal{R}e(H_+)\cos(t)- \mathcal{I}m(H_-)\sin(t)}
=\frac{\mathcal{I}m(H_-)\cos(t)+ \mathcal{R}e(H_+)\sin(t)}
{\mathcal{R}e(H_-)\cos(t)- \mathcal{I}m(H_+)\sin(t)},\label{eqn:tandelta}
\end{eqnarray}

giving $\tan(2t)= \mathcal{I}m\left(2H_+H_-^*\right)/\left(|H_+|^2-|H_-|^2\right)$. 
Then $\delta = \arg(H_+')=\arg(H_-')$ can be determined. 
It is emphasised that in Eqn.~\ref{eqn:tandelta}, $H_+$ and $H_-$ can be chosen to belong to two separate resonances.

The discussions above are also valid for \texttt{2F2P} decays with resonances in all three chains. In this case, four $H$ matrices, built from all possible helicity couplings, would be required to produce all required $P$ and $Q$ terms in the PDF, as

\begin{equation}
\begin{aligned}
\left(\begin{array}{llllll}
H^{R}_+ &\cdots  & H^{R'}_+ &\cdots& H^{R''}_+&\cdots\\
H^{R}_- &\cdots  & H^{R'}_- &\cdots& H^{R''}_-&\cdots\\
\end{array}\right),\,\,\,\,
\left(\begin{array}{llllll}
H^{R}_+ &\cdots  & H^{R'}_+ &\cdots& H^{R''}_-&\cdots\\
H^{R}_- &\cdots  & H^{R'}_- &\cdots& H^{R''}_+&\cdots\\
\end{array}\right),\\
\left(\begin{array}{llllll}
H^{R}_+ &\cdots  & H^{R'}_- &\cdots& H^{R''}_+&\cdots\\
H^{R}_- &\cdots  & H^{R'}_+ &\cdots& H^{R''}_-&\cdots\\
\end{array}\right),\,\,\,\,
\left(\begin{array}{llllll}
H^{R}_+ &\cdots  & H^{R'}_- &\cdots& H^{R''}_-&\cdots\\
H^{R}_- &\cdots  & H^{R'}_+ &\cdots& H^{R''}_+&\cdots\\
\end{array}\right),
\end{aligned}
\label{eqn:H2ThreeChain}
\end{equation}

where subamplitude components $R$, $R'$ and $R''$ belong to the three different chains.

Now it is proved that, in the generic case, a positive  and a negative helicity coupling can be set to be real for \texttt{2F2P} decays. Limited to only one decay chain, two more coupling parameters can be fixed to zero. With the matrix transformation method, No more additional freedoms are found to constrain more helicity parameters.  Since any two helicity couplings can be fixed to be real, helicity couplings are no longer good physical observables to be compared with theoretical calculations or cross-experiments. Instead, the fit fraction (FF) of the $R$ resonant component, defined as~\cite{Dai:2023zms}
\begin{equation}
    \text{FF}_{R}=\frac{\int \text{PDF}(H_+^R,H_-^R, H_+^{R'}=H_-^{R'}=0\text{ for } R'\neq R)d\Phi}{\int \text{PDF}(H_+^R,H_-^R, H_+^{R'},H_-^{R'})d\Phi},\label{eqn:FF}
\end{equation}
only depends on those $P$ and $Q$ parameters, and is thus a quantity independent of conventions on helicity couplings.

\section{Pseudo-experiment results}
Pseudo-experiment studies are performed to numerically verify the conclusions reached in the previous section.  Events are generated for the $B^+\to p \overline{p}^- \pi^+$ decay composed of a set of resonances, and are then studied with amplitude fits considering the aforementioned strategies to fix helicity couplings. The fit is carried out with the unbinned maximum likelihood method, such that the PDF normalisation can be used to fix a helicity coupling parameter, in addition to those interesting ones discussed in the previous section. The parameters that maximise the likelihood determine the results of the fit. Iminuit package~\cite{iminuit} is used to maximise the likelihood with respect to free coupling parameters.  In the following, $\mathcal{R}e(H_+^{R_1})=1$ is always set to $0$ to comply with the PDF normalisation requirement.
The $B^+\to \Delta^{++}(p\pi^+) \overline{p}^-$ decay is taken as the reference channel.

\subsection{One decay chain with multiple  resonances}
\label{sec:one_decay_chain}
Here only the reference channel is considered for the $B^+\to p \overline{p}^- \pi^+$ decay (\textit{i.e.} only a single decay chain), with three $\Delta^{++}$ resonances in the decay $\Delta^{++}\to p\pi^+$. The masses, widths, and $J^\eta$ of these $\Delta^{++}$ states are  listed in the first few rows of  Tab.~\ref{table:toy}. A random number  of about 15000 decays are generated, and the helicity couplings for these resonant contributions are randomly assigned.

\begin{table}
\caption{Resonances considered to  generate  $B^+\to p \overline{p}^- \pi^+$ decays.}
\label{table:toy}       
\centering
\begin{tabular}{l|ccc}
\hline\noalign{\smallskip}
 Resonances     & $J^\eta$      & Mass ($\text{MeV}/c^2$) & Width (\text{MeV})\\ 
\noalign{\smallskip}\hline\noalign{\smallskip}
$\Delta(1600)^{++}$ & $3/2^+$ & 1570 & 50\\
$\Delta(1940)^{++}$ & $3/2^-$ & 2000 & 400\\
$\Delta(1750)^{++}$ & $1/2^+$ & 1721 & 70\\
$\Delta(1700)^{0}$ & $3/2^-$& 1710 & 300\\
$\Delta(1900)^{0}$ & $1/2^-$ & 1860 & 250\\
\noalign{\smallskip}\hline
\end{tabular}
\end{table}

Two amplitude fits  are performed to the pseudo-data. In the scheme \texttt{A} fit, all helicity couplings are floated apart from $\mathcal{I}m(H_+^{R_1})=0$, while in the scheme \texttt{B}, four parameters are fixed to zero following the strategy in section~\ref{section:MultipleSolution}, $\mathcal{I}m(H_+^{R_1})=\mathcal{R}e(H_-^{R_1})=\mathcal{I}m(H_-^{R_1})=\mathcal{I}m(H_-^{R_2})=0$. $R_1$ and $R_2$ can be arbitrarily chosen among the three resonances. According to our discussions in section~\ref{section:MultipleSolution}, the two fits should lead to the same likelihood, and there should be no more multiple solutions for the scheme \texttt{B} fit.

The logarithm likelihood (LL) and the fit fraction of each component obtained with the two fits are listed in Tab.\ref{table:fffit1}. A detailed table of the fit fraction of each component and the interference between each two components are shown in Appendix.\ref{sec:inter}. It is seen that the two fits give identical LL and fit fractions (interferences) up to a numeric precision required by the fitter.  Both fit fractions are also consistent with inputs. It confirms that indeed four helicity parameters can be fixed to reach the same physical result for a single-chained \texttt{2F2P} decay. 

It is noted that the helicity couplings that maximise the likelihood in fit scheme \texttt{A} depend on the initial values of the free parameters in the fit, namely they converge to different positions in the parameter space of equivalent solutions. Practically, it leads to a difficulty in judging whether the fit gives the correct result and comparing different fit results of the same dataset, in particularly when the likelihood is complicated by local minima. Additionally, scheme \texttt{A} encounters challenges in calculating the error matrix using the usual second partial derivative method as in Iminuit, since the second partial derivative matrix is not invertible. While in general no such problems exist for \texttt{B} fit. The  helicity couplings used to  generate pesudo-data and those obtained from the fit using scheme \texttt{B} are shown in Table.\ref{table:inputoutput}, where the uncertainties are reported by the fitter. The fit results are consistent with  input values.

\begin{table}
\caption{The logarithm likelihood and the fit fraction of each component for two schemes of fitting  to a single-chained \texttt{2F2P} decay. In scheme \texttt{A}, $\mathcal{I}m(H_+^{R_1})=0$ is fixed, while in scheme \texttt{B}, $\mathcal{R}e(H_-^{R_1})=\mathcal{I}m(H_-^{R_1})=\mathcal{I}m(H_-^{R_2})=0$ are additionally required, where $R_1=\Delta(1600)^{++}, R_2=\Delta(1750)^{++}$.}
\label{table:fffit1}       
\centering
\begin{tabular}{l|l|llllll}
\hline\noalign{\smallskip}
\multirow{2}{*}{Fit}&\multirow{2}{*}{LL}&\multicolumn{3}{c}{FF}\\\cline{3-5}
& & $\Delta(1600)^{++}$ & $\Delta(1750)^{++}$  & $\Delta(1940)^{++}$ \\
\noalign{\smallskip}\hline\noalign{\smallskip}
\texttt{A}    &  $29823.04$  &   0.484655     &   0.367544  &  0.148588 \\
\texttt{B}    &  $29823.04$  &  0.484653     &   0.367547 &   0.148588  \\
\noalign{\smallskip}\hline
\end{tabular}
\end{table}

\begin{table}
\caption{Helicity couplings used to generate pesudo-data and  results of the fit using scheme \texttt{B}.}
\label{table:inputoutput}       
\centering
\begin{tabular}{c|cc}
\hline\noalign{\smallskip}
Coupling & Input              & Fit result         \\ 
\noalign{\smallskip}\hline\noalign{\smallskip}
$\mathcal{R}e(H_{\Delta(1940)}^{+})$  &  $\phantom-1.51$  & $ \phantom-1.54  \pm  0.06 $\\
$\mathcal{I}e(H_{\Delta(1940)}^{+})$  &  $\phantom-0.35$  & $\phantom-0.37  \pm  0.11 $\\
$\mathcal{I}e(H_{\Delta(1940)}^{-})$  &  $-0.78$  & $ -0.77  \pm  0.08 $\\
$\mathcal{R}e(H_{\Delta(1750)}^{+})$  &  $\phantom-0.32$  & $ \phantom-0.32  \pm  0.02 $\\
$\mathcal{I}e(H_{\Delta(1750)}^{+})$  &  $-0.09$  & $ -0.10  \pm  0.03 $\\
$\mathcal{R}e(H_{\Delta(1750)}^{-})$  &  $\phantom-0.52$  & $ \phantom-0.50  \pm  0.05 $\\
$\mathcal{I}e(H_{\Delta(1750)}^{-})$  &  $\phantom-0.40$  & $ \phantom-0.41  \pm  0.08 $\\
\noalign{\smallskip}\hline
\end{tabular}
\end{table}


To further verify our conclusion, the helicity couplings obtained from scheme \verb|A| fit with an arbitrary set of initial parameters are transformed using the strategy mentioned in Sec.\ref{section:onechannel}. The matrix after transformation is compared with the matrix obtained from the fit in scheme \verb|B|. The matrix  $H_{\verb|A|}$ obtained from scheme \verb|A| fit and the corresponding unitary transformation matrix $U$ are shown in Eqns.\ref{eqn:matra} and \ref{eqn:matru}. The transformed matrix of $H_{\verb|A|}$, $H_{\verb|A|_{trans}}$, and the 
$H_{\verb|B|}$ matrix obtained from scheme \verb|B| fit are shown in Eqns.\ref{eqn:matrat} and \ref{eqn:matrb} respectively. The latter two matrices are identical within the numeric precision of the fitter, as they should be according to the discussion in Sec.\ref{section:onechannel}.

\begin{equation}
    H_{\verb|A|}=
    \left(\begin{array}{ccc}
1 & -3.5264-5.1065i & -3.9552+6.4977i\\ 
5.0599-8.0252i & -0.1755+3.1207i & -11.2024+9.9042i\\ 
\end{array}\right)
\label{eqn:matra}
\end{equation}

\begin{equation}
    U=\left(\begin{array}{cc}
1 & 0\\ 
0 & -0.7839-0.6208i\\ 
\end{array}\right)
\times
    \left(\begin{array}{cc}
1 & 5.0599+8.0252i\\ 
-5.0599+8.0252i & 1\\ 
\end{array}\right)
\label{eqn:matru}
\end{equation}

    \begin{equation}
    H_{\verb|A|_{trans}}=
    \left(\begin{array}{ccc}
1 & -0.3237+0.1019i & -1.5396-0.3657i\\ 
0  & -0.5006-0.4057i & 0.7669i\\ 
\end{array}\right)
\label{eqn:matrat}
\end{equation}

\begin{equation}
    H_{\verb|B|}=
   \left(\begin{array}{ccc}
1 & -0.3236+0.1019i & -1.5397-0.3657i\\ 
0 & -0.5007-0.4057i & 0.7669i\\ 
\end{array}\right)
\label{eqn:matrb}
\end{equation}

\subsection{Multiple decay channels}
\label{sec:multi_decay_chain}
In this study, two $\Delta^0\to\overline{p}\pi$ resonances, listed in the second half of Tab.~\ref{table:toy}, are included in the pseudo-data as a second chain. A total of about 15000 decays is simulated. The $\Delta(1600)^{++}$ resonance is taken as the reference resonance ($i.e.$ $R_1$). To test our fitting strategy obtained in section~\ref{section:MultipleSolution}, the pseudo-data are fitted with three schemes. In scheme $\texttt{A}'$, $\mathcal{I}m(H_+^{R_1})=0$ is set, which is expected to have multiple solutions. In scheme $\texttt{B}'$, $\mathcal{I}m(H_+^{R_1})=\mathcal{I}m(H_-^{R_1})=0$ is fixed, which complies with our strategy of fixing multiple solutions. In scheme $\texttt{C}'$, one additional parameter, $\mathcal{R}e(H_-^{R_1})=0$, is fixed on top of scheme $\texttt{B}'$, which is expected to have a reduced fit quality due to insufficient amount of free parameters.  

The LL and fit fractions for the fits in the three different schemes are listed in Tab.\ref{table:fffitmulti}. A detailed table of fit fractions of each component and interference of each two components are shown in Appendix.\ref{sec:inter}. The LL and fit fraction results for $\texttt{A}'$ and $\texttt{B}'$ are identical within the numeric precision of the fitter. The fit fractions are also consistent with those calculated using input helicity coupling parameters. The error matrix of fit $\texttt{A}'$ can't be properly calculated by the Iminuit fit package. The fit quality of $\texttt{C}'$ is worse than those of $\texttt{A}'$ and $\texttt{B}'$ as can be seen from the smaller LL.


\begin{table}
\caption{The logarithm likelihood and the fit fraction of each resonant component for the three different schemes of fits to two-chained decays.  In scheme $\texttt{A}'$, $\mathcal{I}m(H_+^{R_1})$ is fixed to zero, and in scheme $\texttt{B}'$ $\mathcal{I}m(H_+^{R_1})=\mathcal{I}m(H_-^{R_1})=0$ are required, while in scheme $\texttt{C}'$, $\mathcal{R}e(H_-^{R_1})=0$ is additionally required compared to scheme $\texttt{B}'$, where $R_1=\Delta(1600)^{++}$.}
\label{table:fffitmulti}       
\centering
\begin{tabular}{l|l|llllll}
\hline\noalign{\smallskip}
\multirow{2}{*}{Fit}&\multirow{2}{*}{LL}&\multicolumn{6}{c}{FF}\\\cline{3-8}
& & $\Delta(1600)^{++}$ & $\Delta(1700)^{0}$ & $\Delta(1750)^{++}$ & $\Delta(1900)^{0}$ & $\Delta(1940)^{++}$ & Sum\\
\noalign{\smallskip}\hline\noalign{\smallskip}
 $\texttt{A}'$    &  24210.79   &  0.274907   &  0.051978 & 0.433754   &   0.16862   &   0.060224   & 0.989483  \\
 $\texttt{B}'$    &  24210.79   &  0.274909   &  0.051976 & 0.433758   &   0.16862   &   0.060221   & 0.989484   \\
 $\texttt{C}'$    &  24194.10   &  0.275794   &  0.054087 & 0.435634   &   0.17215   &   0.055602   & 0.993271   \\
\noalign{\smallskip}\hline
\end{tabular}
\end{table}

\begin{figure}
\centering
\resizebox{0.6\textwidth}{!}{%
  \includegraphics{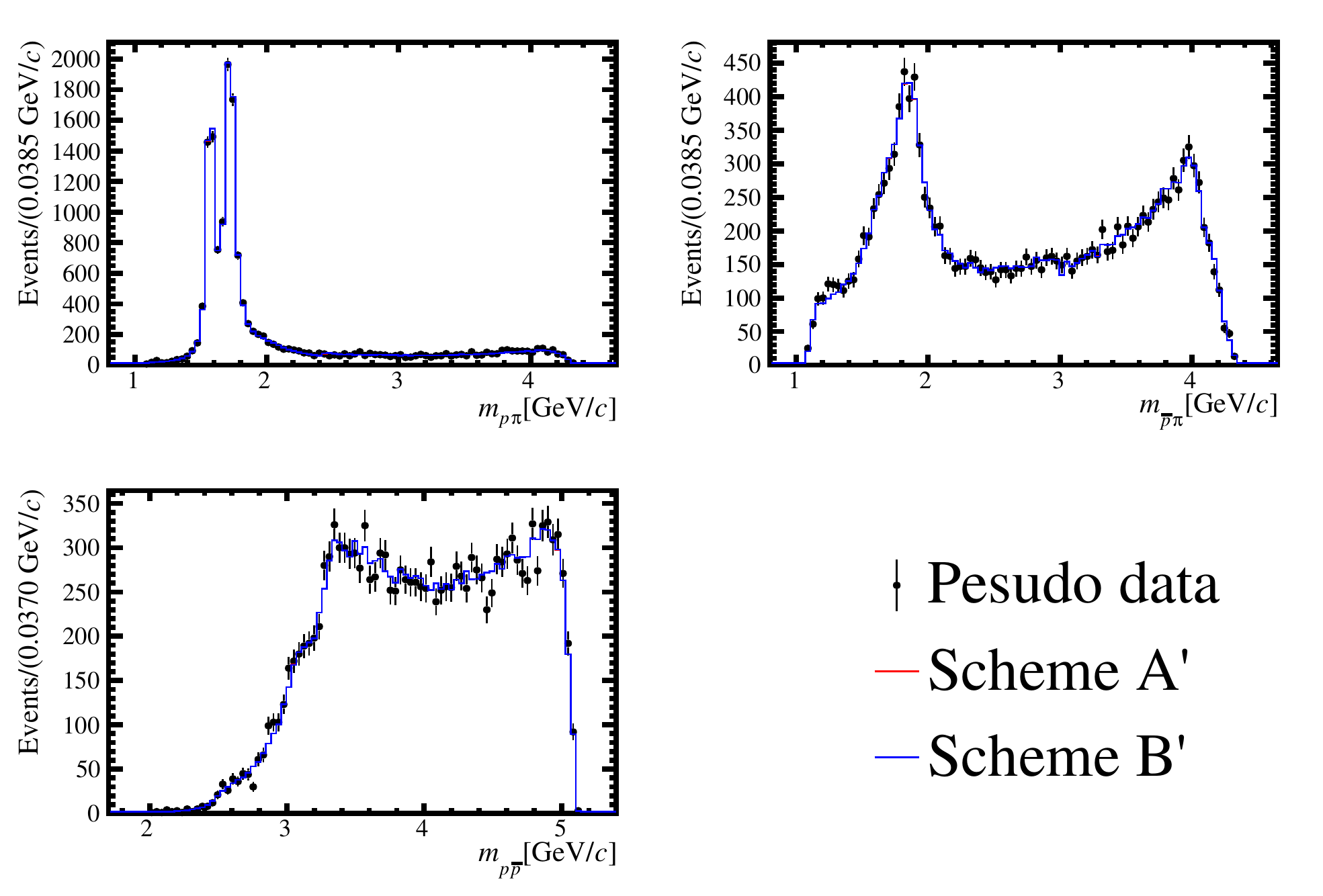}
}
\caption{One-dimensional invariant mass distributions in pseudo-data superimposed by the results in the two fit schemes.}
\label{fig:massdistribution}       
\end{figure}

In Fig.~\ref{fig:massdistribution}, the one-dimensional invariant mass distributions of pseudo-data are shown, overlaid by the results of fit scheme $\texttt{A}'$  and $\texttt{B}'$, showing reasonable consistency between the two fit schemes. And similarly, matrices of helicity couplings obtained from both $\texttt{A}'$ and $\texttt{B}'$ fits, and the transformation matrices are shown in Appendix.\ref{sec:trans_matrix}.

\section{Summaries and conclusions}
Weak decays involving two spin-$1/2$ fermions and two (pseudo-)scalar particles are very common in heavy flavor physics, and are used to study hadron spectroscopy or decay properties, with the aim of searching for new hadrons or testing the Standard Model.
In amplitude fits such decays, the problem of multiple solutions to the helicity couplings is usually encountered, which is studied in detail in this analysis. Detailed studies in this article show that multiple solutions arise since only combinations of helicity couplings are present in the probability density function rather than each coupling independently. 
With multiple solutions, the converged position in the parameter space depends on the initial values of the free parameters, making it difficult to judge whether a fit result is the correct answer or not. Besides, the error matrix of the free parameters cannot be obtained properly using the usual second partial derivative method.

A strategy, obtained using a matrix transformation method, is proposed in our study to resolve the multiple solutions, by properly fixing some helicity couplings. In particular, it is found that if all intermediate resonances contribute to a single decay chain, four helicity parameters can be fixed to zero without any loss of fit quality, three of them belonging to a reference resonance, and another one for a second resonance. If there are multiple decay chains, it is proved that one can set both the positive and negative helicity couplings of a reference resonance to be real, without reducing the fit quality. With the proposed scheme, there is in general no multiple solution problem and the error matrix of free parameters can be calculated from second partial derivatives of the likelihood.
Pseudo-experiments are generated with one or two decay chains to verify the conclusions of the study. 
In this analysis, a strategy is also proposed to align the helicity states of initial- and final-state particles defined in two different decay chains, which compares the coordinate systems defining these helicity states. The results in this study will benefit relevant amplitude analyses at  LHC, BESIII, and B-factory experiments.

\begin{acknowledgement}
This research was funded by the National Natural Science Foundation of China (NSFC) under Contract Nos. 12061141007 and 12175005 and by the National Key R\&D Program of China under Contract No. 2022YFA1601904. 
\end{acknowledgement}

\appendix
\section{Determination of Euler angles}
\label{app:EulerAngles}
Given two coordinate systems $C^0=(\hat{x}^0,\hat{y}^0,\hat{z}^0)$ and 
$C^1=(\hat{x}^1,\hat{y}^1,\hat{z}^1)$, the $C^1$ system is reached from the $C^0$ system by first a rotation of $\alpha$ about the $\hat{z}^0$ axis, followed by a rotation of $\beta$ about the new $\hat{y}'$ axis and then a rotation of $\gamma$ about the new $\hat{z}''$ axis. The three Euler angles are calculated to be 
\begin{eqnarray}
\alpha &=& \texttt{atan2}(\hat{z}^1\cdot\hat{y}^0,\hat{z}^1\cdot\hat{x}^0)\\
\beta &=& \texttt{acos}(\hat{z}^1\cdot\hat{z}^0)\\
\gamma &=& \texttt{atan2}(\hat{z}^0\cdot\hat{y}^1,-\hat{z}^0\cdot\hat{x}^1),
\end{eqnarray}
with the domains  $\beta\in[0,\pi]$, $\alpha,\gamma\in(-\pi,\pi]$ . In literature, the $(\phi,\theta,\psi)\equiv(\alpha,\beta,\gamma)$ notation is also often used.  

\section{Matrix transformation for a single chain}
\label{eqn:PSingleChain}
As  mentioned in section~\ref{section:onechannel}, a $H$ matrix defined as Eqn.~\ref{eqn:H} can be used to determine all the helicity-coupling combinations and the $U$ matrice defined in Eqn.~\ref{eqn:rotation} helps to find a definite solution among multiple solutions. In this part, the detailed calculations are shown.  Without the loss of generality, only three resonances, in the same decay chain, are  considered for simplicity of writing. Now Eqn.~\ref{eqn:H}  turns into

\begin{equation}
H=\left(\begin{array}{llll}
A^{R_1}\eta^{R_1}+iB^{R_1}\eta^{R_1} & A^{R_2}\eta^{R_2}+iB^{R_2}\eta^{R_2}& A^{R_3}\eta^{R_3}+iB^{R_3}\eta^{R_3} \\
C^{R_1}+iD^{R_1} & C^{R_2}+iD^{R_2} & C^{R_3}+iD^{R_3} 
\end{array}\right),
\label{eqn:H3}
\end{equation}

where  the parity symmetry coefficient $\eta$ have been explicitly spelled.
The $U$ matrix in Eqn.~\ref{eqn:rotation} is divided into the product of two parts as $U=U_1\times U_2$, where
\begin{equation}
U_1=\left(\begin{array}{cc}
p_1-ip_2 & p_3-ip_4 \\
-p_3-ip_4 & p_1+ip_2 
\end{array}\right),\; U_2=\left(\begin{array}{cc}
1 & 0\\
0 & \exp^{i\psi}
\end{array}\right),
\label{eqn:rotation2}
\end{equation}
with definitions $p_1=A^{R_1}\eta^{R_1}/\mathcal{N}$, $p_2=B^{R_1}\eta^{R_1}/\mathcal{N}$, $p_3=C^{R_1}/\mathcal{N}$ and $p_4=D^{R_1}/\mathcal{N}$, and\\ $\mathcal{N}=\sqrt{\left(A^{R_1}\right)^2+\left(B^{R_1}\right)^2+\left(C^{R_1}\right)^2+\left(D^{R_1}\right)^2}$. It is noted that $p_1^2+p_2^2+p_3^2+p_4^2=1$.

It is easy to obtain the $U_1$-transformed  $H$ matrix as
\begin{equation}
    \begin{aligned}
    H'\equiv U_1H=\left(
\begin{array}{ccc}
 A^{'R_1} & A^{'R_2}+ iB^{'R_2}  & A^{'R_3}+iB^{'R_3} \\
 0 & C^{'R_2}+iD^{'R_2} &  C^{'R_3}+iD^{'R_3}\\
\end{array}
\right),
    \end{aligned}
\end{equation}
with

    \begin{equation}
    \begin{aligned}
    A^{'R_1}&=\left(A^{R_1}\right)^2+\left(B^{R_1}\right)^2+\left(C^{R_1}\right)^2+\left(D^{R_1}\right)^2,\\ A^{'R_2}&=A^{R_1}A^{R_2}\eta^{R_1}\eta^{R_2}+B^{R_1}B^{R_2}\eta^{R_1}\eta^{R_2}+C^{R_1}C^{R_2}+D^{R_1}D^{R_2},\\ B^{'R_2}&=A^{R_1}B^{R_2}\eta^{R_1}\eta^{R_2}-A^{R_2}B^{R_1}\eta^{R_1}\eta^{R_2}+C^{R_1}D^{R_2}-C^{R_2}D^{R_1},\\ 
    A^{'R_3}&=A^{R_1}A^{R_3}\eta^{R_1}\eta^{R_3}+B^{R_1}B^{R_3}\eta^{R_1}\eta^{R_3}-C^{R_1}C^{R_3}-D^{R_1}D^{R_3},\\ B^{'R_3}&=A^{R_1}B^{R_3}\eta^{R_1}\eta^{R_3}-A^{R_3}B^{R_1}\eta^{R_1}\eta^{R_3}-C^{R_1}D^{R_3}+C^{R_3}D^{R_1},\\
    C^{'R_2}&=A^{R_1}C^{R_2}\eta^{R_1}-A^{R_2}C^{R_1}\eta^{R_2}-B^{R_1}D^{R_2}\eta^{R_1}+B^{R_2}D^{R_1}\eta^{R_2},\\ D^{'R_2}&=A^{R_1}D^{R_2}\eta^{R_1}-A^{R_2}D^{R_1}\eta^{R_2}+B^{R_1}C^{R_2}\eta^{R_1}-B^{R_2}C^{R_1}\eta^{R_2},\\C^{'R_3}&=A^{R_1}C^{R_3}\eta^{R_1}+A^{R_3}C^{R_1}\eta^{R_3}-B^{R_1}D^{R_3}\eta^{R_1}-B^{R_3}D^{R_1}\eta^{R_3},\\ D^{'R_3}&=A^{R_1}D^{R_3}\eta^{R_1}+A^{R_3}D^{R_1}\eta^{R_3}+B^{R_1}C^{R_3}\eta^{R_1}+B^{R_3}C^{R_1}\eta^{R_3}.
    \end{aligned}
    \label{eqn:definition}
\end{equation}

The matrix $H'$ multiplied by $U_2$ becomes
\begin{equation}
    \begin{aligned}
    H^{''}\equiv U_2H'=\left(
\begin{array}{ccc}
 A^{''R_1} &  A^{''R_2}+ iB^{''R_2}  &  A^{''R_3}+ iB^{''R_3}  \\
 0 &  C^{''R_2}+ iD^{''R_2}    & C^{''R_3}+i D^{''R_3}   \\
\end{array}
\right),
    \end{aligned}
\end{equation}
with
\begin{equation}
\begin{aligned}
    A^{''R_1}&=A^{'R_1},\\
    A^{''R_2}&=A^{'R_2},\\
    B^{''R_2}&=B^{'R_2},\\
    A^{''R_3}&=A^{'R_3},\\
    B^{''R_3}&=B^{'R_3},\\
    C^{''R_2}&=C^{'R_2} \cos (\psi)-D^{'R_2} \sin (\psi),\\
    D^{''R_2}&=C^{'R_2} \sin (\psi)+D^{'R_2} \cos (\psi),\\
    C^{''R_3}&=C^{'R_3} \cos (\psi)-D^{'R_3} \sin (\psi),\\
    D^{''R_3}&=C^{'R_3} \sin (\psi)+D^{'R_3} \cos (\psi).\\
\end{aligned}
\label{eqn:AfterU2}
\end{equation}

Therefore, for $\psi=-\arctan(D^{'R_2}/C^{'R_2})+n\pi$, $H^{''}$  turns  into
\begin{equation}
    \begin{aligned}
    H^{''}=U_2U_1H=\left(
\begin{array}{ccc}
 A^{''R_1} & A^{''R_2}+ iB^{''R_2}  & A^{''R_3}+iB^{''R_3} \\
 0 & C^{''R_2} &  C^{''R_3}+iD^{''R_3}\\
\end{array}
\right).
    \end{aligned}
\end{equation}
Namely the four free parameters of a $2\times2$ $U(2)$ matrix have been used to eliminate $\mathcal{I}m(H^{R_1}_+)$, $\mathcal{R}e(H^{R_1}_-)$, $\mathcal{I}m(H^{R_1}_-)$ and $\mathcal{I}m(H^{R_2}_-)$. Of course,  they can be alternatively used to remove other helicity-coupling parameters. For example, according to Eqn.~\ref{eqn:AfterU2}, $\tan(\psi)=C^{'R_2}/D^{'R_2}$ would remove $C^{''R_2}$. In Eqn.~\ref{eqn:H},  the parity symmetry coefficients $\eta^{R_m}$ can also be attached to the $H_-^{R_m}$ couplings ({\em i.e.} $C^{R_m}$ and $D^{R_m}$) rather than $H_+^{R_m}$ couplings, which would eventually yield the same result.

\section{Combinations of helicity couplings in PDF of two-chain decays}
\label{appendix:multi}
For simplicity of writing, a \texttt{2F2P} decay with a total of two resonances in two channels is considered as an example to demonstrate the study, and the intermediate resonances are taken to be fermions. For decays involving both fermionic and bosonic intermediate resonances, the conclusion is the same.
The PDF of this decay has the form of

\begin{equation}
    \begin{aligned}
    \text{PDF}=& \sum _{m _{b},m _{p}}\left|H_{m_b}^{R_1} h_{m_p}^{R_1} d_{m_b m_p}^{J^{R_1}} F^{R_1}
    +\sum _{\lambda_b,\lambda_p} D_{m_b \lambda_b}^{1 / 2} D_{m_p \lambda_p}^{1 / 2} H_{\lambda_b}^{R_2} h_{\lambda_p}^{R_2} d_{\lambda_b \lambda_p}^{J^{R_2}} F^{R_2}\right|^2\\
    =&\underbrace{\sum _{m _{b},m _{p}}\left|H_{m_b}^{R_1} h_{m_p}^{R_1} d_{m_b m_p}^{J^{R_1}} F^{R_1}\right|^2}_{T_1}
    +\underbrace{\sum _{m _{b},m _{p}}\left|\sum_{\lambda_b,\lambda_p} D_{m_b \lambda_b}^{1 / 2} D_{m_p \lambda_p}^{1 / 2} H_{\lambda_b}^{R_2} h_{\lambda_p}^{R_2} d_{\lambda_b \lambda_p}^{J^{R_2}} F^{R_2}\right|^2}_{T_2}\\
    &+\underbrace{2 \sum _{m _{b},m _{p}}\mathcal{R}e\left[\left(H_{m_b}^{R_1} h_{m_p}^{R_1} d_{m_b m_p}^{J^{R_1}} F^{R_1}\right)^* \sum_{\lambda_b,\lambda_p} D_{m_b \lambda_b}^{1 / 2} D_{m_p \lambda_p}^{1 / 2} H_{\lambda_b}^{R_2} h_{\lambda_p}^{R_2} d_{\lambda_b \lambda_p}^{J^{R_2}} F^{R_2}\right]}_{T_3},
\end{aligned}
\label{eqn:twoRtwoC}
\end{equation}
where the two resonances $R_1$ and $R_2$ belong to two separate chains, and $F^{R_1}$ and $F^{R_2}$ are their respective mass lineshapes. Note that Wigner rotations are needed for both particles $b$ and $p$  in the second chain to align their helicities defined in the second chain to those in the first chain. The $H$ (parity violating) and $h$  (parity conserving) parameters are the corresponding helicity couplings for $P_0\to R_{ij} P_k$ and $R_{ij}\to P_i P_j$ decays respectively.

There are three different terms in Eqn.~\ref{eqn:twoRtwoC}, labeled as $T_1$, $T_2$, and $T_3$ respectively, which will be dealt with separately below. The term $T_1$ is simply rewritten as
\begin{equation}
    \begin{aligned}
        &\left|H_{m_b}^{R_1} h_{m_p}^{R_1} d_{m_b m_p}^{J^{R_1}} F^{R_1}\right|^2\\
    =&\left|F^{R_1}\right|^2\left(\left|H_{+}^{R_1}\right|^2\left[\left(d_{++}^{J^{R_1}}\right)^2+\left(d_{+-}^{J^{R_1}}\right)^2\right]+\left|H_{-}^{R_1}\right|^2\left[\left(d_{-+}^{J^{R_1}}\right)^2+\left(d_{--}^{J^{R_1}}\right)^2\right]\right)\\
    =&\left|F^{R_1}\right|^2\left[\left(d_{++}^{J^{R_1}}\right)^2+\left(d_{+-}^{J^{R_1}}\right)^2\right]\left(\left|H_{+}^{R_1}\right|^2+\left|H_{-}^{R_1}\right|^2\right), \\
    \end{aligned}
    \end{equation}
where the trivial helicity couplings  $h^{R_1}_{m_p}$ are omitted. This term gives the $P_{1\cdot1}$ combination of helicity couplings.

The term $T_2$ is expanded to
\begin{equation}
    \begin{aligned}
        &\sum _{m _{b},m _{p}}\left|\sum_{\lambda_b,\lambda_p} D_{m_b \lambda_b}^{1 / 2} D_{m_p \lambda_p}^{1 / 2} H_{\lambda_b}^{R_2} h_{\lambda_p}^{R_2} d_{\lambda_b \lambda_p}^{J^{R_2}} F^{R_2}\right|^2\\
        =&\left|F^{R_2}\right|^2\left[\sum_{\lambda_b \lambda_p \lambda_b^{\prime} \lambda_p^{\prime}} \sum_{m_b m_p}\left(D_{m_b \lambda_b}^{1/2} D_{m_p \lambda_p}^{1/2} H_{\lambda_b}^{R_2} h_{\lambda_p}^{R_2} d_{\lambda_b \lambda_p}^{J^{R_2}}\right)^*\left(D_{m_b \lambda_b^{\prime}}^{1/2} D_{m_p \lambda_p^{\prime}}^{1/2} H_{\lambda_b^{\prime}}^{R_2} h_{\lambda_p^{\prime}}^{R_2} d_{\lambda_b^{\prime} \lambda_p^{\prime}}^{J^{R_2}}\right)\right] \\
        =&\left|F^{R_2}\right|^2\left[\sum_{\lambda_b \lambda_p}\left|H_{\lambda_b}^{R_2} h_{\lambda_p}^{R_2} d_{\lambda_b \lambda_p}^{J^{R_2}}\right|^2\right]\\
        =&\left|F^{R_2}\right|^2\left[\left(d_{++}^{J^{R_2}}\right)^2+\left(d_{+-}^{J^{R_2}}\right)^2\right]\left(\left|H_{+}^{R_2}\right|^2+\left|H_{-}^{R_2}\right|^2\right), 
    \end{aligned}
\end{equation}
where  the unitarity of  Wigner-D functions
\begin{equation}
    \sum_k D_{m^{\prime} k}^j(\alpha, \beta, \gamma)^* D_{m k}^j(\alpha, \beta, \gamma)=\delta_{m, m^{\prime}}
\end{equation}
 has been used to obtain the third equation. Again the trivial helicity couplings  $h^{R_2}_{\lambda_p}$ are omitted. This term gives the $P_{2\cdot2}$ term.

The term $T_3$ is expanded as
\begin{equation}
    \begin{aligned}
    &2 \sum_{m_b m_p}\left(H_{m_b}^{R_1} h_{m_p}^{R_1} d_{m_b m_p}^{J^{R_1}} F^{R_1}\right)^*\left(\sum_{\lambda_b \lambda_p} D_{m_b \lambda_b}^{1 / 2} D_{m_p \lambda_p}^{1 / 2} H_{\lambda_b}^{R_2} h_{\lambda_p}^{R_2} d_{\lambda_b \lambda_p}^{J^{R_2}}\right) F^{R_2} \\
    =&2 F^{R_1 *} F^{R_2} \sum_{m_b m_p \lambda_b \lambda_p}\left(H_{m_b}^{R_1} h_{m_p}^{R_1} d_{m_b m_p}^{J^{R_1}}\right)^*\left(D_{m_b \lambda_b}^{1 / 2} D_{m_p \lambda_p}^{1 / 2} H_{\lambda_b}^{R_2} h_{\lambda_p}^{R_2} d_{\lambda_b \lambda_p}^{J^{R_2}}\right) \\
    =&F^{R_1 *} F^{R_2}\sum_{m_b m_p \lambda_b \lambda_p}\left[H_{m_b}^{R_1 *} H_{\lambda_b}^{R_2} h_{m_p}^{R_1} h_{\lambda_p}^{R_2} D_{m_b \lambda_b}^{1 / 2} D_{m_p \lambda_p}^{1 / 2} d_{m_b m_p}^{J^{R_1}} d_{\lambda_b \lambda_p}^{J^{R_2}}\right.\\
    &\hspace{0.22\textwidth}\left.+H_{-m_b}^{R_1 *} H_{-\lambda_b}^2 h_{-m_p}^1 h_{-\lambda_p}^2 D_{-m_b-\lambda_b}^{1 / 2} D_{-m_p-\lambda_p}^{1 / 2} d_{-m_b-m_p}^{J^{R_1}} d_{-\lambda_b-\lambda_p}^{J^{R_2}}\right] \\
    =&F^{R_1 *} F^{R_2} \sum_{m_b m_p \lambda_b \lambda_p}e^ {-i m_b \phi_b-i m_p \phi_p}h_{m_p}^{R_1} h_{\lambda_p}^{R_2}\left[\left(H_{m_b}^{R_1 *} H_{\lambda_b}^{R_2}+\text{S}\times H_{-m_b}^{R_1 *} H_{-\lambda_b}^2\right) d_{m_b \lambda_b}^{1 / 2} d_{m_p \lambda_p}^{1 / 2} d_{m_b m_p}^{J^{R_1}} d_{\lambda_b \lambda_p}^{J^{R_2}}\right],
\end{aligned}
\end{equation}

where the constant $\text{S}\equiv e^ {2i (m_b \phi_b- m_p \phi_p)}\eta^{R_1} \eta^{R_2}$ is either $+1$ or $-1$ depending on the Wigner rotation angles $\phi_b, \phi_p=0 \text{ or }\pi$ and the two parity parameters $\eta^{R_1} \eta^{R_2}$, but not depending on helicities $m_b, m_p$. The second equation is obtained by using the fact that for a term with helicities $\{m_b, m_p, \lambda_b, \lambda_p\}$, there is always a corresponding  one with $\{-m_b, -m_p, -\lambda_b, -\lambda_p\}$. The parity symmetry for helicity couplings and the relation $d^j_{-m',-m}=(-1)^{m-m^\prime}d_{m',m}^j$ for $d$-functions have been used to derive the last (third) equation.
This term $T_3$ gives the helicity coupling combinations in Eqn.~\ref{eqn:PQformulti}.

\section{Fit fractions for simulation results including interference between different resonances}
\label{sec:inter}
The fit fraction of each component and the interference between every two components for simulation in Sec.\ref{sec:one_decay_chain} and Sec.\ref{sec:multi_decay_chain} are shown in this section.
Tables~\ref{tab:FFA} and~\ref{tab:FFB}  summarize the FFs and interferences for the scheme \texttt{A} and scheme \texttt{B} fits data of the single chained $$B^+\to p \overline{p}^- \pi^+$$ decay. Tables~\ref{tab:FFAp} and~\ref{tab:FFBp}  summarize the FFs and interferences for the scheme \texttt{A'} and scheme \texttt{B'} fits data of the two chained $B^+\to p \overline{p}^- \pi^+$ decay. Refer to the text for a detailed description of the fit schemes.

\begin{table}
\caption{Fit fractions obtained using scheme $\texttt{A}$ fit. The off-diagonal components are for interferences. Due to the symmetry of the off-diagonal components, only the upper part is displayed while the lower part is replaced with '--'}
\label{tab:FFA}       
\centering
\begin{tabular}{cccc}
\hline\noalign{\smallskip}
                     & $\Delta(1600)^{++}$ & $\Delta(1750)^{++}$ & $\Delta(1940)^{++}$   \\
\noalign{\smallskip}\hline\noalign{\smallskip}
$\Delta(1600)^{++}$  & 0.484655            & $-$0.000420            & 0.000139     \\
$\Delta(1750)^{++}$  & --                  & 0.367544               & $-$0.000507  \\
$\Delta(1940)^{++}$  & --                  & --                     & 0.148588     \\
\noalign{\smallskip}\hline
\end{tabular}
\end{table}

\begin{table}
\caption{Fit fractions obtained using scheme $\texttt{B}$ fit. The off-diagonal components are for interferences. Due to the symmetry of the off-diagonal components, only the upper part is displayed while the lower part is replaced with '--'}
\label{tab:FFB}       
\centering
\begin{tabular}{cccc}
\hline\noalign{\smallskip}
                     & $\Delta(1600)^{++}$ & $\Delta(1750)^{++}$ & $\Delta(1940)^{++}$   \\
\noalign{\smallskip}\hline\noalign{\smallskip}
$\Delta(1600)^{++}$  & 0.484653            & $-$0.000420            & 0.000139            \\
$\Delta(1750)^{++}$  & --            & 0.367547            & $-$0.000507           \\
$\Delta(1940)^{++}$ & --            & --           & 0.148588                 \\
\noalign{\smallskip}\hline
\end{tabular}
\end{table}

\begin{table}
\caption{Fit fractions obtained using scheme $\texttt{A}^{\prime}$ fit. The off-diagonal components are for interferences. Due to the symmetry of the off-diagonal components, only the upper part is displayed while the lower part is replaced with '--'}
\label{tab:FFAp}       
\centering
\begin{tabular}{cccccc}
\hline\noalign{\smallskip}
& $\Delta(1600)^{++}$ & $\Delta(1700)^{++}$ & $\Delta(1750)^{++}$ & $\Delta(1900)^{++}$ & $\Delta(1940)^{++}$   \\
\noalign{\smallskip}\hline\noalign{\smallskip}
$\Delta(1600)^{++}$ & 0.274907            & 0.000591            & $-$0.000407           & $-$0.004418           & 0.007996            \\
$\Delta(1700)^{++}$ & --            & 0.051978           & 0.021664            & 0.001391            & 0.001788           \\
$\Delta(1750)^{++}$ & --           & --            & 0.433754            & 0.017107            & $-$0.000605           \\
$\Delta(1900)^{++}$ & --           & --            & --            & 0.168620             & $-$0.018600             \\
$\Delta(1940)^{++}$ & --            & --           & --           & --             & 0.060224            \\
\noalign{\smallskip}\hline
\end{tabular}
\end{table}

\begin{table}
\caption{Fit fractions obtained using scheme $\texttt{B}^{\prime}$ fit. The off-diagonal components are for interferences. Due to the symmetry of the off-diagonal components, only the upper part is displayed while the lower part is replaced with '--'} 
\label{tab:FFBp}       
\centering
\begin{tabular}{cccccc}
\hline\noalign{\smallskip}
& $\Delta(1600)^{++}$ & $\Delta(1700)^{++}$ & $\Delta(1750)^{++}$ & $\Delta(1900)^{++}$ & $\Delta(1940)^{++}$   \\
\noalign{\smallskip}\hline\noalign{\smallskip}
$\Delta(1600)^{++}$ & 0.274909            & 0.000593            & $-$0.000407           & $-$0.004418           & 0.008002            \\
$\Delta(1700)^{++}$ & --            & 0.051976           & 0.021663            & 0.001391            & 0.001787           \\
$\Delta(1750)^{++}$ & --           & --            & 0.433758            & 0.017109            & $-$0.000605           \\
$\Delta(1900)^{++}$ & --           & --            & --            & 0.168620             & $-$0.018600             \\
$\Delta(1940)^{++}$ & --            & --           & --           & --             & 0.060221            \\
\noalign{\smallskip}\hline
\end{tabular}
\end{table}

\section{Detailed helicity couplings for multiple decay channels}
\label{sec:trans_matrix}

Matrices of helicity couplings obtained from a fit to pseudo-data of two chains of the $B^+\to p \overline{p}^- \pi^+$ decay. The matrix $H_{\verb|A|^{\prime}}$ is formed by helicity couplings obtained from scheme $\verb|A|^{\prime}$ fit with an arbitrary set of initial
parameters, which doesn't converge with correct error matrix. The matrix $H_{\verb|B|^{\prime}}$ is built from helicity couplings obtained from scheme $\verb|B|^{\prime}$ fit, which converges properly. The matrix $U$ is a unitary transformation that brings the helicity couplings of the reference resonance to be real, $i.e.$ that of Eqn.~\ref{eqn:UmatrixSpecial}. Under the transformation of $U$, the matrix $H_{\verb|A|^{\prime}}$ becomes $H_{\verb|A|^{\prime}_{trans}}$, which is compared to the matrix $H_{\verb|B|^{\prime}}$.
\begin{equation}
H_{\verb|A|^{\prime}}=\left(\begin{array}{ccccc}
1+0i & -1.9333-2.1460i & -3.6880-1.9885i & 0.9339+2.2177i & -2.6738+1.4067i\\
2.9673-0.7981i & 0.9111+0.81522i & 0.3518+1.3642i & -0.0967-2.3994i & 2.2107+1.2875i\\
\end{array}\right)
\end{equation}

\begin{equation}
        H_{\verb|B|^{\prime}}=\left(\begin{array}{ccccc}
1+0i & -1.0658-2.8354i & -2.810-3.1818i & -0.0919+2.4206i & -2.8802+0.2964i\\
2.9672+0i & 0.3124+1.2448i & -0.4720+1.8034i & 0.9904-2.4216i & 1.8090+2.3633i\\
\end{array}\right)
\end{equation}

\begin{equation}
H_{\verb|A|^{\prime}_{trans}}=\left(\begin{array}{ccccc}
1+0i & -1.0656-2.830542i & -2.8088-3.1761i & -0.0914+2.4174i & -2.8747+0.2985i\\
2.9629+0i & 0.3130+1.2441i & -0.4699+1.8008i & 0.9868-2.4189i & 1.8088+2.3600i\\
\end{array}\right)
\end{equation}

\begin{equation}
U=(0.9473+0.3203i)\times\left(\begin{array}{cc}
0.1132+0i & 0.9935i\\
0.9935i & 0.1132+0i\\
\end{array}\right)
\end{equation}


%
\bibliographystyle{unsrt}
\bibliography{sample}

\begin{thebibliography}{10}

\bibitem{CMS:2012vby}
Serguei Chatrchyan et~al.
\newblock {Study of the Mass and Spin-Parity of the Higgs Boson Candidate Via
  Its Decays to Z Boson Pairs}.
\newblock {\em Phys. Rev. Lett.}, 110(8):081803, 2013.

\bibitem{LHCb:2015yax}
Roel Aaij et~al.
\newblock {Observation of $J/\psi p$ Resonances Consistent with Pentaquark
  States in $\Lambda_b^0 \to J/\psi K^- p$ Decays}.
\newblock {\em Phys. Rev. Lett.}, 115:072001, 2015.

\bibitem{LHCb:2022nyw}
{Direct $CP$ violation in charmless three-body decays of $B^{\pm}$ mesons}.
\newblock 6 2022.

\bibitem{Fleming:1964zz}
Gordon~N. Fleming.
\newblock {Recoupling Effects in the Isobar Model. 1. General Formalism for
  Three-Pion Scattering}.
\newblock {\em Phys. Rev.}, 135:B551--B560, 1964.

\bibitem{osti_4551044}
D~Morgan.
\newblock Phenomenological analysis of ${I} = 1/2$ single-pion production
  processes in the energy range 500 to 700 {MeV}.
\newblock {\em Phys. Rev., 166: 1731-59(Feb. 25, 1968).}, 1 1968.

\bibitem{PhysRevD.11.3165}
David~J. Herndon, Paul S\"oding, and Roger~J. Cashmore.
\newblock Generalized isobar model formalism.
\newblock {\em Phys. Rev. D}, 11:3165--3182, Jun 1975.

\bibitem{LHCb:2017jym}
Roel Aaij et~al.
\newblock {Study of the $D^0 p$ amplitude in $\Lambda_b^0\to D^0 p \pi^-$
  decays}.
\newblock {\em JHEP}, 05:030, 2017.

\bibitem{Zhang:2021sit}
Shunan Zhang, Yi~Jiang, Zewen Chen, and Wenbin Qian.
\newblock {Sensitivity studies on the CKM angle $\gamma$ in $\Lambda_b^0 \to
  D\Lambda$ decays}.
\newblock 12 2021.

\bibitem{LHCb:2021enr}
Roel Aaij et~al.
\newblock {Search for $CP$ violation in $\Xi^-_b \to p K^- K^-$decays}.
\newblock {\em Phys. Rev. D}, 104(5):052010, 2021.

\bibitem{LHCb:2022ouv}
{Amplitude analysis of the $\Lambda^+_c\to pK^-\pi^+$ decay and $\Lambda^+_c$
  baryon polarization measurement in semileptonic beauty hadron decays}.
\newblock 8 2022.

\bibitem{Zemach:1965ycj}
Charles Zemach.
\newblock {Use of angular momentum tensors}.
\newblock {\em Phys. Rev.}, 140:B97--B108, 1965.

\bibitem{Jacob:1959at}
M.~Jacob and G.~C. Wick.
\newblock {On the General Theory of Collisions for Particles with Spin}.
\newblock {\em Annals Phys.}, 7:404--428, 1959.

\bibitem{PDG}
R.~L. Workman et~al.
\newblock {Review of Particle Physics}.
\newblock {\em PTEP}, 2022:083C01, 2022.

\bibitem{Chen:2017gtx}
Hong Chen and Rong-Gang Ping.
\newblock {Coherent helicity amplitude for sequential decays}.
\newblock {\em Phys. Rev. D}, 95(7):076010, 2017.

\bibitem{JPAC:2019ufm}
M.~Mikhasenko et~al.
\newblock {Dalitz-plot decomposition for three-body decays}.
\newblock {\em Phys. Rev. D}, 101(3):034033, 2020.

\bibitem{Marangotto:2019ucc}
Daniele Marangotto.
\newblock {Helicity Amplitudes for Generic Multibody Particle Decays Featuring
  Multiple Decay Chains}.
\newblock {\em Adv. High Energy Phys.}, 2020:6674595, 2020.

\bibitem{Wang:2020giv}
Mengzhen Wang, Yi~Jiang, Yinrui Liu, Wenbin Qian, Xiaorui Lyu, and Liming
  Zhang.
\newblock {A novel method to test particle ordering and final state alignment
  in helicity formalism}.
\newblock {\em Chin. Phys. C}, 45(6):063103, 2021.

\bibitem{Dai:2023zms}
Xinchen Dai, Miroslav Saur, Yiduo Shang, Xueting Yang, and Yanxi Zhang.
\newblock {CP Violation in Baryon Decays at LHCb}.
\newblock {\em Symmetry}, 15(2):522, 1 2023.

\bibitem{iminuit}
Hans Dembinski and Piti~Ongmongkolkul et~al.
\newblock scikit-hep/iminuit.
\newblock Dec 2020.

\end{thebibliography}

\end{document}